\documentclass[12pt]{article}
\usepackage[utf8]{inputenc}
\usepackage{amsmath, amssymb}
\usepackage{braket}
\usepackage{ascmac}
\usepackage{color}
\usepackage{here}
\usepackage{slashed}
\usepackage{url}
\usepackage{mathrsfs}
\usepackage[colorlinks=true, linkcolor=black, citecolor=blue, urlcolor=blue]{hyperref}
\usepackage{graphicx}
\usepackage[symbol]{footmisc}

\numberwithin{equation}{section}

\usepackage[top=25mm, bottom=30mm, left=30mm, right=30mm]{geometry}

\newcommand{\nn}{\nonumber}
\newcommand{\del}{\partial}

\usepackage{ulem} 
\usepackage{color}

\begin{document}

\thispagestyle{empty}

\begin{flushright}
    \small UTCCS-P-156,
    \small KANAZAWA 24-07, 
    \small RIKEN-iTHEMS-Report-24
\end{flushright}

\begin{center}
{\Large \textbf{
Grassmann tensor renormalization group for the massive Schwinger model with a $\theta$ term using staggered fermions
}}

\vspace*{5mm}
\textsc{Hayato Kanno}$^{a,}$\footnote[1]{
    \texttt{\href{mailto:hayato.kanno@riken.jp}{hayato.kanno@riken.jp}}
}, 
\textsc{Shinichiro Akiyama}$^{b,c,}$\footnote[2]{
    \texttt{\href{mailto:akiyama@ccs.tsukuba.ac.jp}{akiyama@ccs.tsukuba.ac.jp}}
},\\
\textsc{Kotaro Murakami}$^{d,e,}$\footnote[3]{
\texttt{\href{mailto:kotaro.murakami@yukawa.kyoto-u.ac.jp}{kotaro.murakami@yukawa.kyoto-u.ac.jp}}
}, \textsc{Shinji Takeda}$^{f,}$\footnote[4]{
\texttt{\href{mailto:takeda@hep.s.kanazawa-u.ac.jp}{takeda@hep.s.kanazawa-u.ac.jp}}
}

\vspace*{4mm}

\hspace{-0.5cm}
{\fontsize{11pt}{12pt}\selectfont
\textit{{$^a$
RIKEN BNL Research Center,
Brookhaven National Laboratory, Upton, NY 11973, USA 
}}\\
\vspace*{1mm}
\textit{{$^b$
Center for Computational Sciences,
University of Tsukuba, Tsukuba, Ibaraki 305-8577, Japan
}}\\
\vspace*{1mm}
\textit{{$^c$
Graduate School of Science,
The University of Tokyo, Bunkyo-ku, Tokyo 113-0033, Japan
}}\\
\vspace*{1mm}
\textit{{$^d$
Department of Physics, Institute of Science Tokyo, 2-12-1 Ookayama, Megro, Tokyo 152-8551, Japan
}}\\
\vspace*{1mm}
\textit{{$^e$
RIKEN Center for Interdisciplinary Theoretical and Mathematical Sciences (iTHEMS), RIKEN, Wako, Saitama 351-0198, Japan
}}\\
\vspace*{1mm}
\textit{{$^f$
Institute for Theoretical Physics, Kanazawa University, Kanazawa 920-1192, Japan
}}\\
}
\end{center}

\vspace*{1mm}

\begin{abstract}
We use the Grassmann tensor renormalization group method to investigate the $N_f=2$ Schwinger model with the staggered fermions in the presence of a $2\pi$ periodic $\theta$ term in a broad range of mass.
The method allows us to deal with the massive staggered fermions straightforwardly and to study the $\theta$ dependence of the free energy and topological charge in the thermodynamic limit. 
Our calculation provides consistent results with not only the analytical solution in the large mass limit but also the previous Monte Carlo studies in the small mass regime.
Our numerical results also suggest that the $N_f=2$ Schwinger model on a lattice has a different phase structure, than the model in the continuum limit.
\end{abstract}

\newpage
\setcounter{page}{1}
\setcounter{footnote}{0}
\renewcommand{\thefootnote}{\arabic{footnote}}

\tableofcontents

\section{Introduction}

Understanding the topological nature of Quantum ChromoDynamics (QCD) is one of the essential subjects in high-energy physics. 
There is a famous unsolved problem in the standard model called the strong CP problem, which is the unnaturalness that the $\theta$ term in the QCD action almost vanishes according to the neutron EDM experiments~\cite{Abel:2020pzs}. 
One of the candidates to explain such a phenomenon is the Peccei-Quinn mechanism~\cite{Peccei:1977hh}, where the $\theta$ parameter behaves as a dynamical field, which is called the axion field. 
The studies on the axion have attracted much attention so far, as a solution not only to the strong CP problem itself but also to other phenomenological puzzles such as dark matter~\cite{Preskill:1982cy, Abbott:1982af, Dine:1982ah} and inflation~\cite{Freese:1990rb}.
On the other hand, to determine the properties of the axion, non-perturbative studies on QCD with the $\theta$ term are vital. 
For studies on QCD without $\theta$ term, numerical simulation of lattice gauge theories using the Monte Carlo (MC) technique is a powerful tool. 
The MC technique, however, does not work when including the $\theta$ term since the Boltzmann weight in the path integral is a complex number.
Such difficulty is called the sign problem. 
Although there are some attempts to study the four-dimensional (4D) Yang-Mills theory with the $\theta$ term~\cite{Kitano:2021jho, Yamada:2024vsk, Hirasawa:2024fjt}, the MC approach to the lattice QCD studies with the $\theta$ term have not been developed so much.

Recently, numerical techniques using tensor networks have become prominent candidates for avoiding the problem. 
The most conventional one is the Density Matrix Renormalization Group (DMRG)~\cite{White:1992zz} based on the Hamilton formalism~\cite{Kogut:1974ag}. 
Another numerical approach is the tensor renormalization group (TRG)~\cite{Levin:2006jai}, in which we represent the path integral in the Lagrangian formalism as a tensor network.
However, application to the four-dimensional (4D) QCD is still challenging for both approaches since their computational cost is large. 
Under these circumstances, in this paper, we numerically study the Schwinger model~\cite{Schwinger:1962tp}, the 2D Quantum ElectroDynamics (QED), with the $\theta$ term. 
The model is well known as a toy model of the 4D QCD and has a similar infrared (IR) phase structure to the 4D QCD since they share similar global symmetry and confinement nature. 
Moreover, the Schwinger model has a nontrivial topological nature related to the $\theta$ term as well as 4D QCD.

In this study, we consider the two-flavor ($N_f=2$) Schwinger model.
In the previous studies, the one-flavor Schwinger model had been mainly studied. 
In spite of this, it is also worth exploring the phenomena in the multi-flavor theory for the understanding of vacuum structure in more realistic situations. 
We emphasize that the vacuum structure of $N_f=1$ and $N_f\ge2$ are quite different at $\theta=\pi$; there is a transition from unique vacuum to two-fold vacuums at a certain mass parameter for $N_f=1$ while two-fold vacuum degeneracy appears at any positive mass for $N_f\ge2$.
In the massless limit, the two-flavor Schwinger model is exactly solvable in the IR limit because it becomes a conformal field theory (CFT): the $SU(N_f)_1$ Wess-Zumino-Witten (WZW) model.
The theory in the heavy mass limit can also be solved because it corresponds to the 2D Maxwell theory ($U(1)$ gauge theory), where no propagating degrees of freedom appear. 
In contrast to these two limits, it is difficult to solve the model analytically in the general finite mass case.
The exception is the small mass case; it is understood well through the bosonized theory with the mass perturbation~\cite{Coleman:1975pw, Coleman:1976uz}

We employ the TRG method to simulate the Schwinger model with the $\theta$ term.
One of the advantages of the TRG approach is that it allows us to handle lattice volumes large enough to be identified as the thermodynamic limit.
The TRG method is also particularly advantageous for the simulation on a torus, which is necessary to realize the $2\pi$ periodicity with respect to the $\theta$ parameter.
In addition, TRG can treat fermionic degrees of freedom directly as Grassmann variables~\cite{Gu:2010yh}, which makes it easy to formulate the tensor network representation even for theories including fermions. 
Taking these advantages, in this paper, we perform the TRG algorithm to calculate the free energy density in the Schwinger model with the $2\pi$ periodic $\theta$ term in a wide range of mass parameters. 
Furthermore, we use single-component staggered fermion action, which corresponds to the two-flavor Schwinger model in the continuum limit. 
This action is suitable for a first trial of the simulation using TRG since the calculation cost is lower than those for other lattice actions such as the two-flavor Wilson fermion.

For the lattice calculation of the Schwinger model with the $\theta$ term, 
several approaches in the MC method have already been taken~\cite{Fukaya:2003ph, Fukaya:2004kp, Azcoiti:2017mxl}.
Simulations on the massless Schwinger model based on the MC method using a dual formulation have also been performed~\cite{Gattringer:2015nea, Goschl:2017kml}, in which the sign problem is avoided.
Remarkably, there are MC simulations for the bosonized Schwinger model with the $\theta$ term~\cite{Ohata:2023sqc, Ohata:2023gru}, where the sign problem is eliminated by integrating out the gauge field in the bosonized action. 
Recently, there have been many studies on the Schwinger model with the $\theta$ term using other numerical methods, such as quantum computation~\cite{Chakraborty:2020uhf,Pederiva:2021tcd,Honda:2021aum,Angelides:2023noe, Kaikov:2024acw, Li:2024jlo} and several Matrix Product State (MPS) methods including the DMRG (see Refs.~\cite{Byrnes:2002gj, Funcke:2019zna, Dempsey:2022nys, Angelides:2023bme, ArguelloCruz:2024xzi, Fujii:2024reh} for $N_f=1$ and Refs.~\cite{Dempsey:2023gib, Itou:2023img, Itou:2024psm} for $N_f=2$, and the references therein).
Furthermore, there have been previous studies by the TRG method using the one-flavor Wilson fermion action~\cite{Shimizu:2014uva, Shimizu:2014fsa, Shimizu:2017onf} and the massless staggered fermion~\cite{Butt:2019uul}. 
Our study is the first TRG calculation of the two-flavor massive Schwinger model with the $\theta$ term.
In particular, we perform the thermodynamic calculations on the free energy and topological charge, which were considered difficult in the previous tensor network formulation proposed in Ref.~\cite{Butt:2019uul}. 

This paper is organized as follows. 
In Section~\ref{sec:review}, we briefly review several known facts for the two-flavor Schwinger model. 
We describe the lattice action for the Schwinger model and summarize its tensor network representation in Section~\ref{sec:lattice}.  
Our numerical results for the $N_{f}=2$ Schwinger model are presented in Section~\ref{sec:results}.
Section~\ref{sec:conclusion} is devoted to the summary.

\section{Review of the Schwinger model}
\label{sec:review}

In this section, we review the analytic results of the Schwinger model which is the 2D gauge theory including a $U(1)$ gauge field (photon) and $N_f$ Dirac fermions in fundamental representation. 
Here, we exclusively consider the case where all the $N_f(\ge 2)$ fermions share the same mass.  
The Euclidean action in this theory is given by
\begin{align}
S =& \int {\rm d}^2x \bigg\{\frac{1}{4g^2}F_{\mu\nu}F^{\mu\nu} + \frac{{\rm i}\theta}{4\pi}\epsilon^{\mu\nu}F_{\mu\nu} + \bar{\psi}{\rm i}\gamma^{\mu}(\del_{\mu}+{\rm i}A_{\mu})\psi + m\bar{\psi}\psi  \bigg\} \ ,
\label{contS}
\end{align}
where $m$($\ge0$) is the mass parameter of the fermions. 
The second term in the above equation corresponds to the $\theta$ term in this model.
Since the Schwinger model has the same properties as those of the 4D QCD, a bunch of theoretical studies in this model have been done as a testbed of QCD.

The analytical studies of the Schwinger model have been done well in the large mass limit, massless limit, and finite but very small mass region.
In the following, we review the analytic results for each mass regime. 

In a large mass limit, the Schwinger model is just reduced to a pure $U(1)$ gauge theory, that is, the 2D Maxwell theory. 
This theory no longer has dynamical particles since the gauge field has no degree of freedom to propagate. 
The free energy, however, has $\theta$-dependence since the $\theta$ term play a role of the background static electric field.

In massless limit, the Schwinger model can be analyzed through its global symmetry.
The massless Schwinger model with $N_f$-flavor has $\frac{U(N_f)_L \times U(N_f)_R}{U(1)_V\times U(1)_A} \simeq SU(N_f)_L \times SU(N_f)_R$ global symmetry. 
Note that $U(1)_V$ is not a global symmetry because it is gauged.
On the other hand, $U(1)_A$ global symmetry is broken because of the ABJ-type anomaly, which is the same case as for 4D QCD.
Although there is no spontaneous symmetry breaking because of the Coleman-Mermin-Wagner theorem\cite{Mermin:1966fe, Coleman:1973ci}, the global symmetry of this model is similar to that of 4D QCD.
As in the case of 4D QCD, the massless theory is independent of $\theta$ since the $\theta$ term is always compensated by the ABJ-type anomaly through a proper $U(1)_A$ transformation.
Furthermore, the massless Schwinger model is similar to 4D QCD in the sense of the infrared (IR) effective theory, even though it does not have spontaneous symmetry breaking.
For 4D QCD, the IR effective theory is known as a $\left(SU(N_f)_L \times SU(N_f)_R\right)/SU(N_f)_V$ non-linear sigma model that comes from the chiral symmetry breaking. This is nothing but the pion effective theory. 
On the other hand, the massless Schwinger model in the IR limit is equivalent to the $SU(N_f)_1$ WZW model, which is a conformal field theory where the central charge is $c=N_f-1$. This relation can be derived by the non-abelian bosonization~\cite{Witten:1983ar, Gepner:1984au}. Since the action of the WZW model is very similar to the pion theory, the massless Schwinger model has a similar IR structure to that of 4D QCD.

For a finite mass parameter, the theory preserves only $SU(N_f)_V$ symmetry, the same global symmetry as in 4D QCD. This theory then can have a non-trivial topological $\theta$ term because of the homotopy group for the $U(1)$ gauge symmetry, $\pi_1(U(1))=\mathbb{Z}$.~\footnote{
    For the 4D QCD, the $\theta$ term comes from $\pi_3(SU(N_c))=\mathbb{Z}$. The 2D QCD and 4D QED cannot have a $\theta$ term on $\mathbb{R}^d$ or $S^d$ because $\pi_1(SU(N))=0$ for the 2D QCD and $\pi_3(U(1))=0$ for the 4D QED.
}
It is known that this theory has a mass gap. 
There is a gapless point at $\theta=\pi$ and a certain point of $m\neq0$ for $N_f=1$, while the theory for $N_f\ge2$ has a mass gap in the whole $m>0$ regime even at $\theta=\pi$~\cite{Dempsey:2023gib}.~\footnote{
    This behavior is the same as the IR effective theory of 4D QCD, in which the gapless point exists  at a $m>0$ and $\theta=\pi$ point for $N_f=1$ and at $m=0$ point for $N_f\ge2$ \cite{Gaiotto:2017tne}. This similarity of the phase diagram is understood through the non-abelian bosonization.
}
In contrast to the large and small mass limits, it is difficult to calculate the Schwinger model analytically for the finite mass in general. 
For this reason, in this paper, we employ a numerical calculation.

In this paper, we mainly focus on the $\theta$ dependence of the free energy densities for $N_f=2$.
Here, we summarize the analytic solution for large and small mass limits. The detailed calculation is given in Appendix~\ref{app:analytic}. 
First, the free energy density in a large mass limit, which is just the one in the Maxwell theory, can be calculated as
\begin{align}
    -\frac{\log Z(\theta)}{g^2 V} =&  \min_n  \frac{1}{8\pi^2}\left(\theta -2\pi n \right)^2 \ .
    \label{eq:f_Maxwell_result}
\end{align}
We normalize it by $g^2$ to make a dimension-less combination. The derivation is shown in Appendix~\ref{app:large}.
Note that the above equation is evaluated in the large volume limit (thermodynamic limit).
Next, for a small mass regime, the free energy density is evaluated by abelian bosonization together with mass perturbation. 
Although the detailed calculation is presented in Appendix~\ref{app:small}, the result of the free energy density from the mass perturbation is given by, \cite{Coleman:1976uz}
\begin{align}
    -\frac{\log Z(\theta)}{g^2 V}
    =& \min_n \left\{\left(\mathrm{e}^{\gamma}\right)^{\frac{4}{3}}\pi^{-\frac{5}{3}}2^{\frac{1}{3}}\left(\frac{m^2}{g^2}\right)^{\frac{2}{3}}\cos^{\frac{4}{3}}\left(\frac{\theta-2\pi n}{2}\right) \right\} \ 
    \label{eq:f_exact_light_result}
\end{align}
with the Euler constant, $\gamma = 0.57721\cdots$.
Note that this equation is valid as long as the mass perturbation works, that is, only for the small $m/g$ parameter range.
There are no free parameters in the factors in Eq.~(\ref{eq:f_Maxwell_result}) and Eq.~(\ref{eq:f_exact_light_result}), so we can compare these values with numerical results.

\section{Tensor network representation of path integral}
\label{sec:lattice}

\subsection{Lattice action}
\label{sec:action}

To simulate the $N_{f}=2$ Schwinger model on a square lattice, we use the standard Wilson gauge action for the $U(1)$ gauge symmetry with the $2\pi$-periodic $\theta$ term and staggered fermion action:
\begin{align}
\label{eq:action}
	S=S_{g}+S_{\Theta}+S_{f}
	,
\end{align}
where
\begin{align}
	S_{g}=
	-\beta\sum_{n\in\Lambda_{2}}
	\Re\left[U_{1}(n)U_{2}(n+\hat{1})U^{*}_{1}(n+\hat{2})U^{*}_{2}(n)\right]
	,
\end{align}
\begin{align}
	S_{\Theta}=
    -\frac{\theta}{2\pi}\sum_{n}
	\log\left[U_{1}(n)U_{2}(n+\hat{1})U^{*}_{1}(n+\hat{2})U^{*}_{2}(n)\right]
	, \label{eq:theta_term}
\end{align}
\begin{align}
	S_{f}=
	\frac{1}{2}\sum_{n}\sum_{\nu=1}^{2}
	\eta_{\nu}(n)\left[\bar{\chi}(n)U_{\nu}(n)\chi(n+\hat{\nu})-\bar{\chi}(n+\hat{\nu})U^{*}_{\nu}(n)\chi(n)\right]
	+m_0\sum_{n}\bar{\chi}(n)\chi(n)
	.
\end{align}
Here, $m_0=am$ and $\beta=1/(ag)^2$ are the mass parameter and inverse gauge coupling with a lattice spacing $a$, respectively.
The $U(1)$-valued link variable is denoted by $U_{\nu}(n)$ and $\chi(n)$ and $\bar{\chi}(n)$ are the single-component Grassmann variables. 
The staggered sign function is defined by $\eta_{1}(n)=1$ and $\eta_{2}(n)=(-1)^{n_{1}}$.
We assume the periodic (anti-periodic) boundary condition for the staggered fermions in $\hat{1}$ ($\hat{2}$) direction.
This action is defined on a 2D Euclidean lattice.
Thus one staggered fermion includes $2^{2}=4$ complex valued fermionic degrees of freedom, and corresponds to $N_f=2$ Dirac fermion in the continuum limit~\cite{Kogut:1974ag}.

In Eq.~\eqref{eq:theta_term}, $\log$ denotes the principal value.
This logarithmic definition of a $\theta$ term enables us to treat integer-valued instanton numbers because the spacetime sum of this topological charge density should result in an integer value~\cite{Seiberg:1984ya}.
The logarithmic definition has been widely applied in studying topological aspects in 2D lattice gauge theories~\cite{Fukaya:2003ph,Fukaya:2004kp,Shimizu:2014fsa,Kawauchi:2016xng,Kuramashi:2019cgs,Hirasawa:2021qvh,Nakayama:2021iyp}.

Parametrizing the link variable as $U_{\nu}(n)={\rm e}^{{\rm i}A_{\nu}(n)}$, the partition function is defined by the path integral as
\begin{align}
\label{eq:lat_path_int}
	Z=
	\left(\prod_{n}\int\int{\rm d}\bar{\chi}(n){\rm d}\chi(n)\right)
	\left(\prod_{n,\nu}\int^{\pi}_{-\pi}\frac{{\rm d}A_{\nu}(n)}{2\pi}\right)
	{\rm e}^{-S}
	.
\end{align}

\subsection{Tensor network formulation}
\label{sec:TRG}

We represent the path integral in Eq.~\eqref{eq:lat_path_int} as a tensor network.
Following Ref.~\cite{Kuramashi:2019cgs}, we discretize the $U(1)$-valued link variables by the Gauss-Legendre quadrature rule.
Firstly, we define a four-leg tensor $T^{(g)}$ as
\begin{align}
\label{eq:gauge_tensor}
	&T^{(g)}_{a_{2}(n+\hat{1})a_{1}(n+\hat{2})a_{2}(n)a_{1}(n)}
	=
	\frac{\sqrt{w_{a_{1}(n)}w_{a_{2}(n+\hat{1})}w_{a_{1}(n+\hat{2})}w_{a_{2}(n)}}}{4}
	\nonumber\\
	&\times
	{\rm e}^{
            \beta\cos\left[
                \pi\left(a_{1}(n)+a_{2}(n+\hat{1})-a_{1}(n+\hat{2})-a_{2}(n)\right)
            \right]
            +
            \frac{\theta}{2\pi}\log\left[
                {\rm e}^{{\rm i}\pi\left(a_{1}(n)+a_{2}(n+\hat{1})-a_{1}(n+\hat{2})-a_{2}(n)\right)}
            \right]
        }
	.
\end{align}
In Eq.~\eqref{eq:gauge_tensor}, $a_{\nu}(n)$ denotes the node of the Gauss-Legendre quadrature rule and $w_{a_{\nu}(n)}$ is the corresponding weight:
\begin{align}
\label{eq:def_gl_quadrature}
	\int^{1}_{-1}
        {\rm d}x_{\nu}(n)
        f\left(x_{\nu}(n)\right)
	\simeq
	\sum_{a_{\nu}(n)\in D_{K}}
        w_{a_{\nu}(n)}
        f\left(a_{\nu}(n)\right),
\end{align}
where $x_{\nu}(n)=A_{\nu}(n)/\pi$ and $f$ represents the corresponding integrand.
$D_{K}$ is a set of $K$ sampling points defined by the quadrature rule.
Note that the efficacy of this quadrature rule has been demonstrated in Ref.~\cite{Kuramashi:2019cgs}, where the first-order transition at $\theta=\pi$ in the 2D $U(1)$ lattice gauge theory is captured even with relatively small $K$.

To deal with the Grassmann integrals, we employ the Grassmann tensor network formulation in Ref.~\cite{Akiyama:2020sfo}.
Introducing the auxiliary Grassmann fields, we can define the Grassmann tensor in the following form,
\begin{align}
\label{eq:fermion_tensor}
    &\mathcal{T}^{(f)}_{
        \zeta_{1}\xi_{1}
        \zeta_{2}\xi_{2}
        \bar{\xi}_{1}\bar{\zeta}_{1}
        \bar{\xi}_{2}\bar{\zeta}_{2}
        ,a_{1}(n)a_{2}(n)
    }
    =
    \prod_{\nu}\sum_{i_{\nu},j_{\nu},i'_{\nu},j'_{\nu}}
    T^{(f)}_{
        i_{1}j_{1}i_{2}j_{2}i'_{1}j'_{1}i'_{2}j'_{2}
        ,a_{1}(n)a_{2}(n)
    }
    \zeta_{1}^{i_{1}}\xi_{1}^{j_{1}}
    \zeta_{2}^{i_{2}}\xi_{2}^{j_{2}}
    \bar{\xi}_{1}^{j'_{1}}\bar{\zeta}_{1}^{i'_{1}}
    \bar{\xi}_{2}^{j'_{2}}\bar{\zeta}_{2}^{i'_{2}}
    .
\end{align}
In Eq.~\eqref{eq:fermion_tensor}, $\zeta_{\nu}$, $\xi_{\nu}$, $\bar{\zeta}_{\nu}$, $\bar{\xi}_{\nu}$ represent the auxiliary single-component Grassmann fields and the bits $i_{\nu}$, $j_{\nu}$, $i'_{\nu}$, $j'_{\nu}$ does the occupation numbers.
The coefficient tensor $T^{(f)}$ depends on the discretized gauge fields and the staggered sign function. 
For simplicity, we have omitted the site dependence from the auxiliary Grassmann fields.
The explicit form of $T^{(f)}$ is derived in Appendix~\ref{app:derivation_tn}.

\begin{figure}[htbp]
    \centering
    \includegraphics[width=1\hsize]{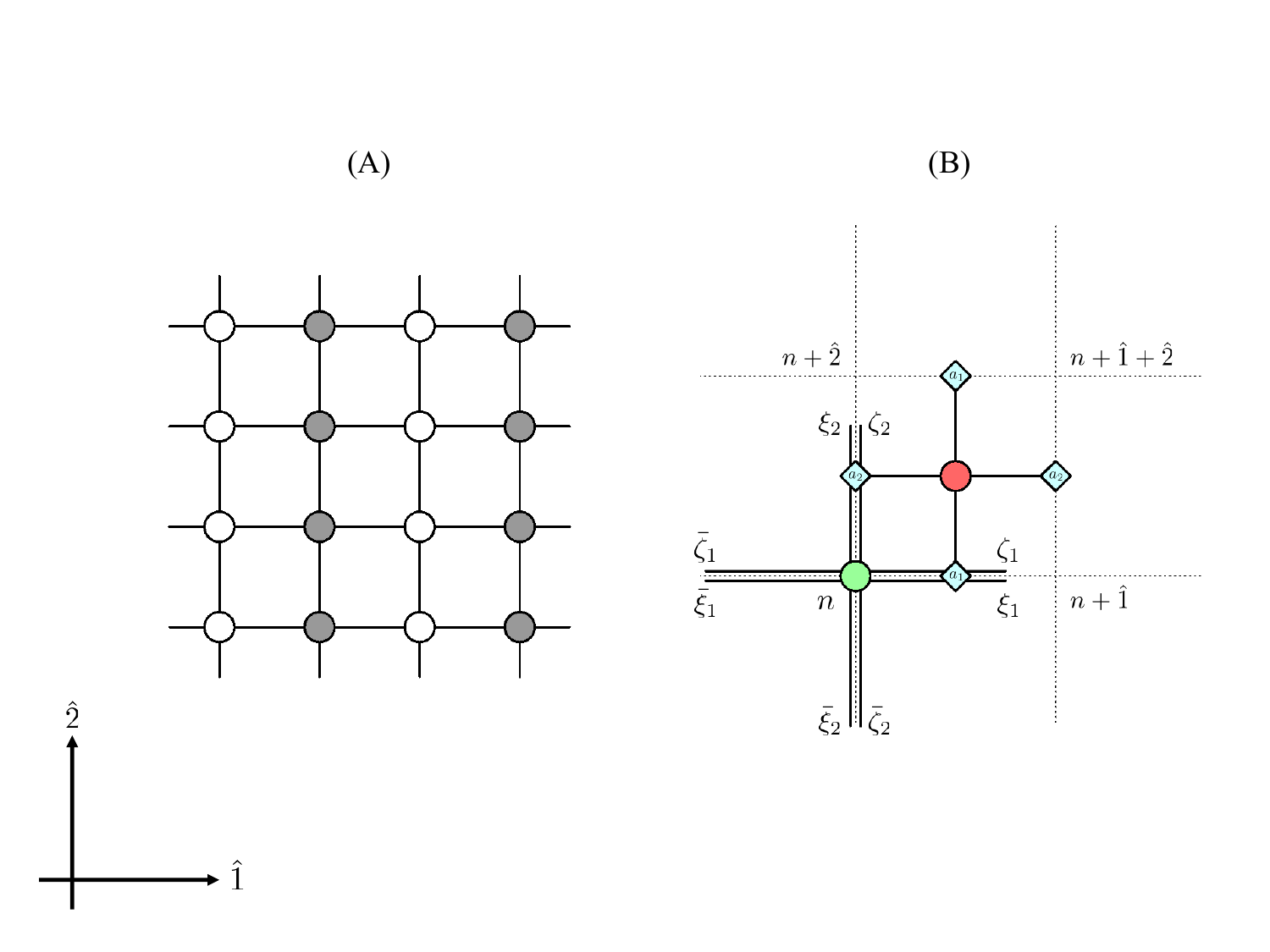}
    \caption{
        (A) Schematic picture of the Grassmann tensor network in Eq.~\eqref{eq:tn_rep}.
        Since the fundamental tensor defined in Eq.~\eqref{eq:fundamental_tensor} depends on $n_{1}$ due to the staggered sign function, two kinds of tensors, white and gray symbols, are necessary to restore the path integral.
        (B) Structure of the fundamental tensor in Eq.~\eqref{eq:fundamental_tensor}.
        Red and green symbols show $T^{(g)}$ and $\mathcal{T}^{(f)}$, respectively.
        Dotted lines represent the square lattice.
        Each external line represents the auxiliary Grassmann field.
        Gauge fields are denoted by the diamonds.
    }
    \label{fig:tn_rep}
\end{figure}

Combining two types of tensors in Eqs.~\eqref{eq:gauge_tensor} and \eqref{eq:fermion_tensor}, we can define a fundamental Grassmann tensor associated with each lattice site $n$ as
\begin{align}
\label{eq:fundamental_tensor}
    \mathcal{T}_{n;
        \zeta_{1}\xi_{1}
        \zeta_{2}\xi_{2}
        \bar{\xi}_{1}\bar{\zeta}_{1}
        \bar{\xi}_{2}\bar{\zeta}_{2}
        ,a_{2}(n+\hat{1})a_{1}(n+\hat{2})a_{2}(n)a_{1}(n)
    }
    =
    T^{(g)}_{a_{2}(n+\hat{1})a_{1}(n+\hat{2})a_{2}(n)a_{1}(n)}
    \mathcal{T}^{(f)}_{
        \zeta_{1}\xi_{1}
        \zeta_{2}\xi_{2}
        \bar{\xi}_{1}\bar{\zeta}_{1}
        \bar{\xi}_{2}\bar{\zeta}_{2}
        ,a_{1}(n)a_{2}(n)
    }
    ,
\end{align}
which describes the original path integral, Eq.~\eqref{eq:lat_path_int}, via
\begin{align}
\label{eq:tn_rep}
    Z\simeq Z(K)
    =
    {\rm gTr}
    \left[
        \prod_{n}
        \mathcal{T}_{n}
    \right].
\end{align}
Notice that ``gTr" does not only mean the integration over the auxiliary Grassmann fields but also the summation over the discretized gauge fields.
Figure~\ref{fig:tn_rep} diagrammatically shows the current tensor network formulation.

The tensor network representation in Eq.~\eqref{eq:tn_rep} is ready to be computed by the TRG method.
The basic idea of the TRG method is to approximately carry out the tensor contraction based on the singular value decomposition (SVD).
The SVD allows us to construct a coarse-grained transformation that best approximates the Frobenius norm of the local fundamental tensor under a fixed tensor size.
This size is usually referred to as the bond dimension.
By increasing the bond dimension, we can systematically improve the accuracy of the TRG method.
Typically, the TRG algorithms allow us to compute a contraction consisting of $2^{n}$ tensors just in $n$ times of coarse-grained transformation.
Therefore, we can easily access the thermodynamic limit.
In this study, we employ the bond-weighted TRG (BTRG) algorithm~\cite{PhysRevB.105.L060402} to evaluate Eq.~\eqref{eq:tn_rep}.
The BTRG improves the accuracy of the original Levin-Nave TRG~\cite{Levin:2006jai} without increasing the computational cost.
This is achieved by introducing some weight on each edge in the tensor network by which the effect of the environment neglected in the original TRG is partly taken into account.
The efficiency of the BTRG for lattice fermions has already been confirmed in Ref.~\cite{Akiyama:2022pse}. 
We always set the hyperparameter $k$ for the weight on each edge as $k=-1/2$, which is the optimal choice in the case of square lattice models.
Our implementation of the BTRG requires the $O(D^{4})$ memory cost and $O(D^{6})$ computational complexity with the bond dimension $D$.
For more information on the BTRG algorithm in the presence of the lattice fermions, see Ref.~\cite{Akiyama:2022pse}, or a recent review paper~\cite{Akiyama:2024ush}.

We finally note that the Schwinger model with the staggered fermions was previously investigated by the TRG method in Ref.~\cite{Butt:2019uul}, where the tensor network representation of the path integral was derived based on the world-line formulation~\cite{Gattringer:2015nea,Goschl:2017kml} and no Grassmann variable appeared.
However, this strategy is highly limited to the massless case as pointed out in Ref.~\cite{Butt:2019uul}.
In contrast, our approach is based on the Grassmann tensor network formulation and there is no difficulty in applying the TRG method even in the presence of the finite fermion mass.
This is a direct benefit because the TRG algorithms can directly deal with the Grassmann variables; the tensor network representation can be constructed using only local fundamental tensors when the original lattice theory is local.

\section{Numerical results}
\label{sec:results}

\subsection{Algorithmic parameters}
\label{subsec:parameters}

We define the dimensionless free energy density by
\begin{align}
    \label{eq:dimeless_f}
    f
    =-\frac{1}{g^2V}\log Z
    =-\frac{\beta}{L^2}\log Z , 
\end{align}    
where $L$ is the linear system size in the lattice unit ($V=a^2L^2$). 
We use the BTRG algorithm to compute Eq.~\eqref{eq:dimeless_f} in the thermodynamic limit.
The physical quantities, including the topological charge and susceptibility, are immediately obtained by taking the numerical difference of Eq.~\eqref{eq:dimeless_f}.

\begin{figure}[htbp]
    \centering
    \includegraphics[width=0.49\hsize]{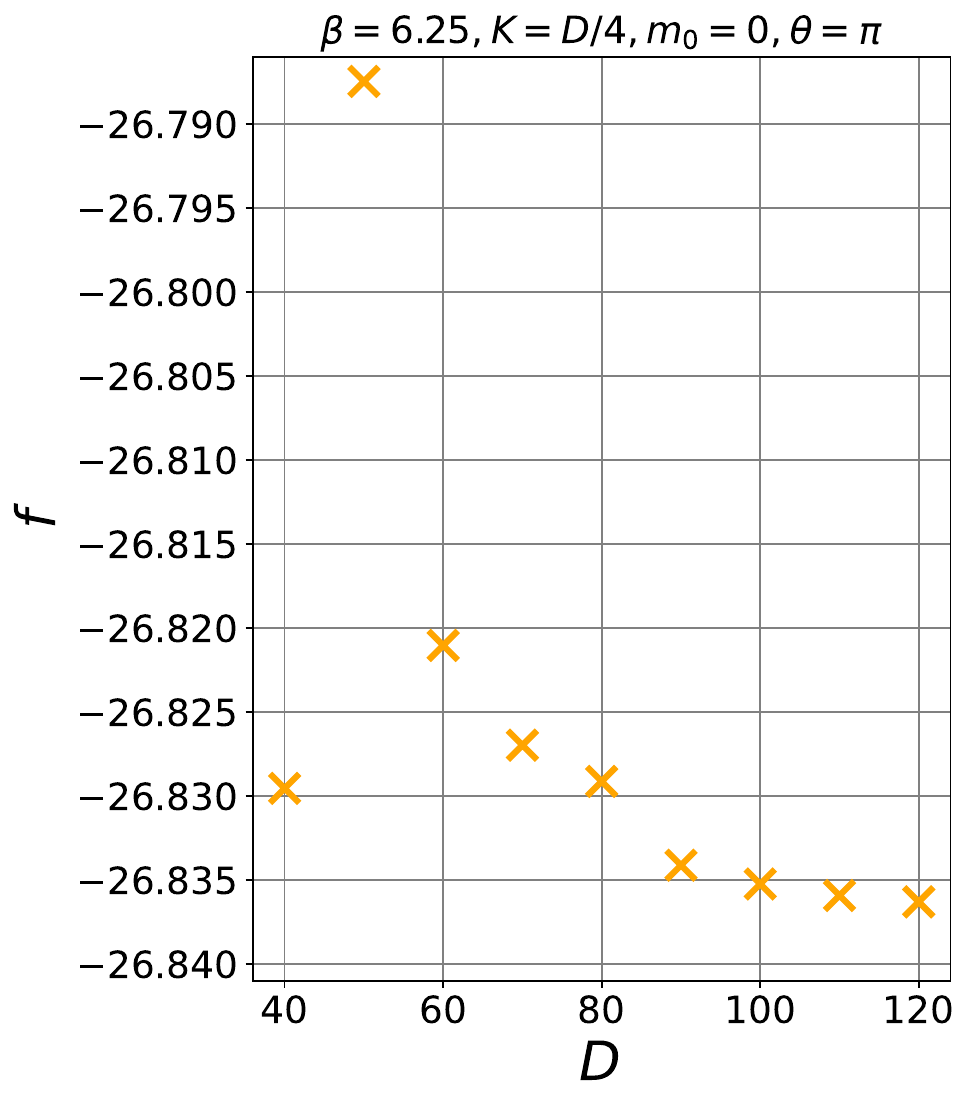}
    \includegraphics[width=0.49\hsize]{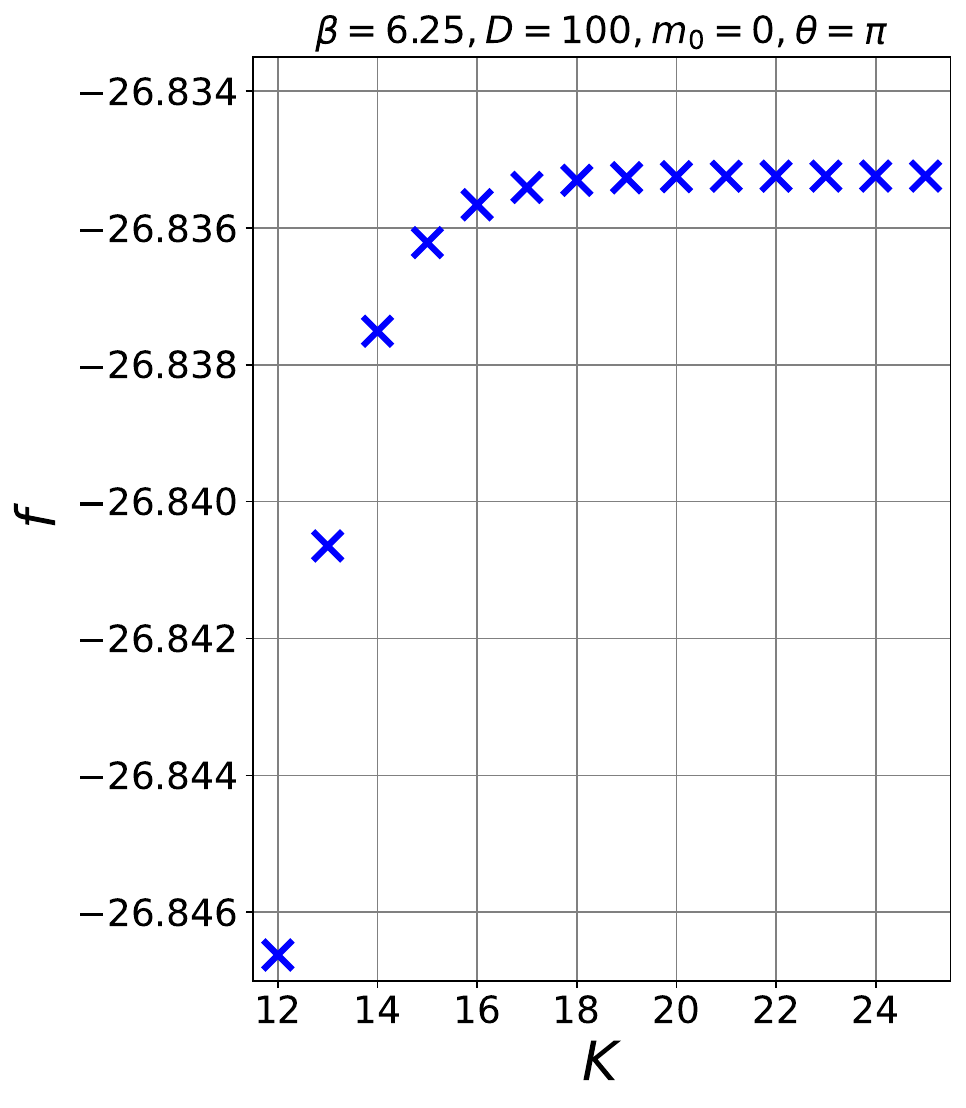}
    \caption{
        Free energy density as a function of the bond dimension $D$ (left) and the cutoff $K$ (right) in the Gauss-Legendre quadrature rule.
    }
    \label{fig:DK}
\end{figure}

We first investigate the convergence of free energy in terms of the cutoff $K$ in the quadrature rule and the bond dimension $D$ in the BTRG algorithm.
As we see below, we employ the inverse gauge coupling with $\beta\le 1/(0.4)^2 =6.25$ in this study.
Here, we set $m_0=0$, $\beta=6.25$, $\theta=\pi$, that requires the largest $K$ and $D$ to reach sufficient convergence.
Fig.~\ref{fig:DK} shows the thermodynamic free energy density as a function of $D$ and $K$.
In the left-hand side of Fig.~\ref{fig:DK}, we set $K=D/4$ because the initial bond dimension of the Grassmann tensor network in Eq.~\eqref{eq:tn_rep} is $4K$ and we are allowed to increase $K$ when the maximal bond dimension $D$ is increased.
In the right-hand side of Fig.~\ref{fig:DK}, the cutoff $K$ is varied with the fixed bond dimension $D=100$.
The finite-$K$ effect seems to be well suppressed with $K\ge14$.
The absolute difference between the resulting free energy with $D=110$ and $D=120$ is about $4\times 10^{-4}$, much smaller than the scale we shall see in the following.
In this study, $D=120$ and $K=25$ are the maximal algorithmic parameters.
In the following, when there is no specific mention of bond dimension or K, we always use $D=120$ and $K=25$.

\subsection{Free energy density}
\label{subsec:free_energy}

\begin{figure}[htbp]
    \centering
    \includegraphics[width=0.94\hsize]{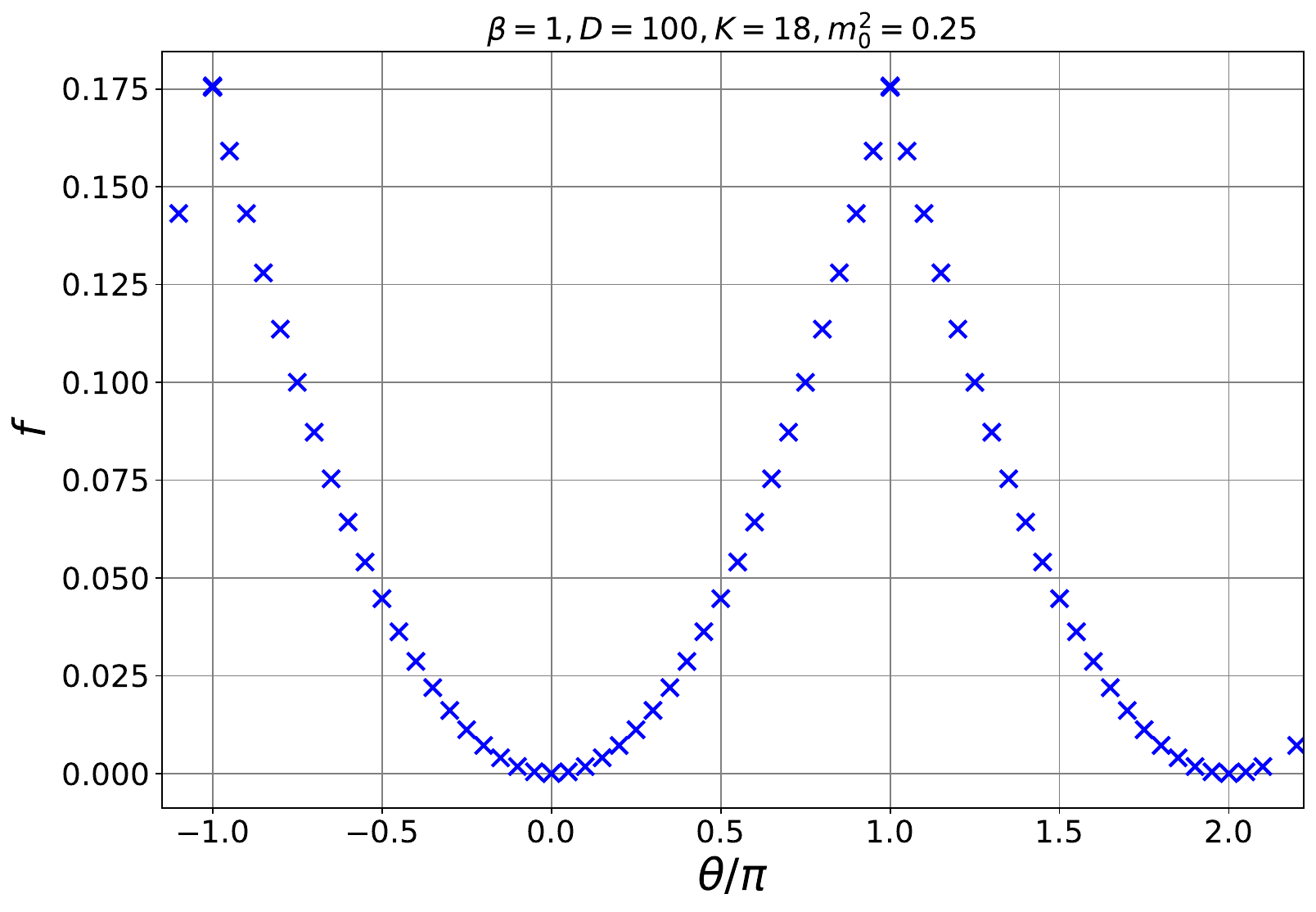}
    \caption{
        Free energy density as a function of the $\theta$, in the range of $-1.1\pi\le\theta\le2.2\pi$.
    }
    \label{fig:wide}
\end{figure}

Let us investigate the free energy density as a function of $\theta$.
In the following, we always consider the shifted free energy via
\begin{align}
\label{eq:shifted_free_energy}
    f(\theta)=
    -\frac{\beta}{L^2}
    \left(  
        \log Z(\theta)
        -
        \log Z(\theta=0)
    \right)
    ,
\end{align}
so that it takes zero at the origin.
Fig.~\ref{fig:wide} shows that the resulting free energy density explicitly has the $2\pi$ periodicity with respect to $\theta$. 
In Fig.~\ref{fig:wide}, we set $\beta=1$ and $m_{0}^{2}=0.25$ ($\sqrt{\beta m_0^2}=0.5$) as a representative.

\begin{figure}[H]
    \centering
    \includegraphics[width=0.48\hsize]{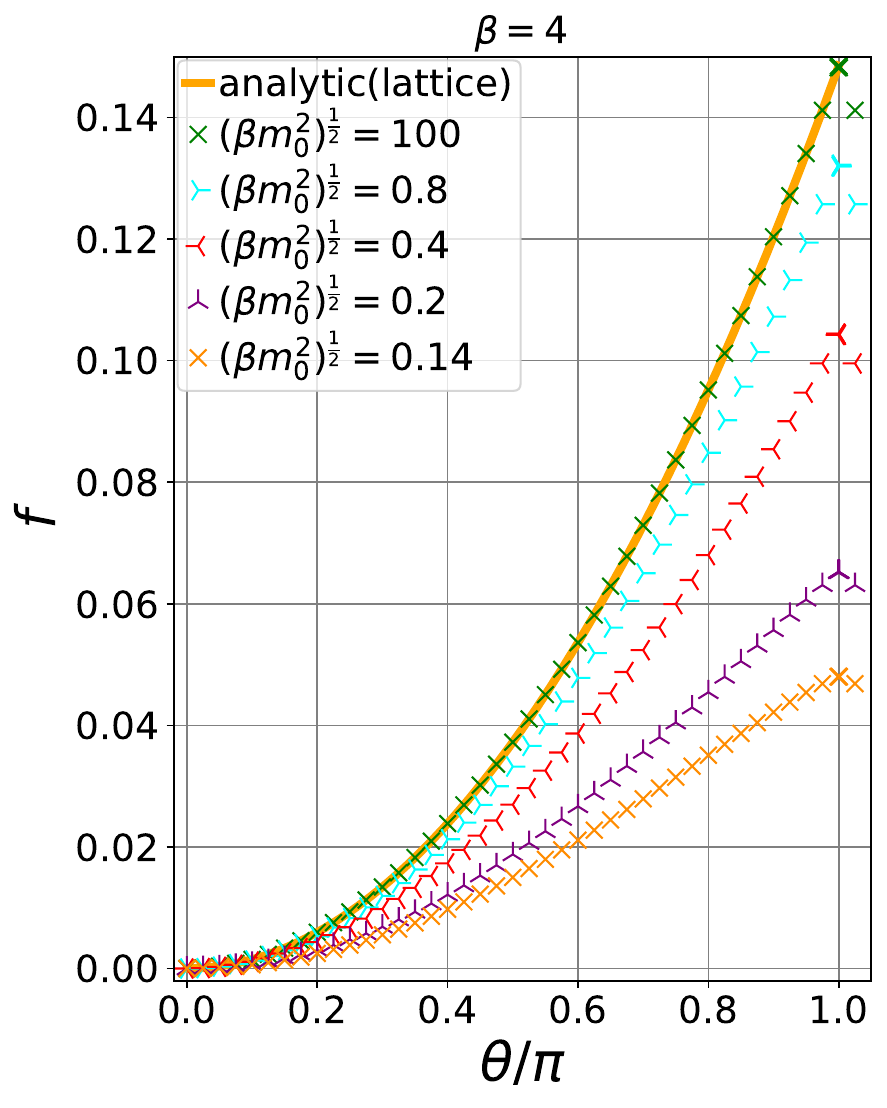}
    \includegraphics[width=0.5\hsize]{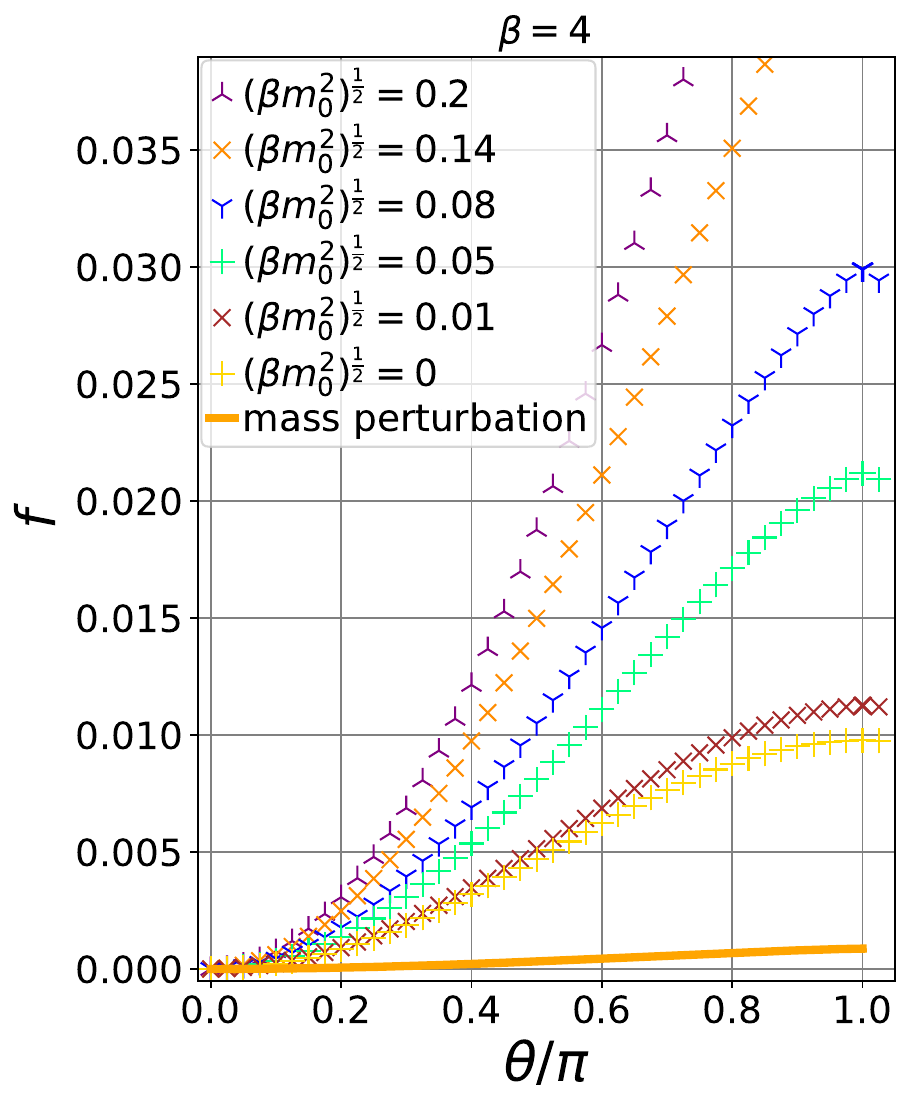}
    \caption{
        Free energy density as a function of $\theta/\pi$ with $\sqrt{\beta m_0^2} \geq 0.14$ (left) and $\sqrt{\beta m_0^2} \leq 0.2$ (right). 
        A solid curve shows the analytical solution of the Maxwell theory on a lattice in the left panel while the mass perturbation result for $\sqrt{\beta m_0^2} =0.01$ in the right.
        }
    \label{fig:mass_f}
\end{figure}

\begin{figure}[H]
    \centering
    \includegraphics[width=0.48\hsize]{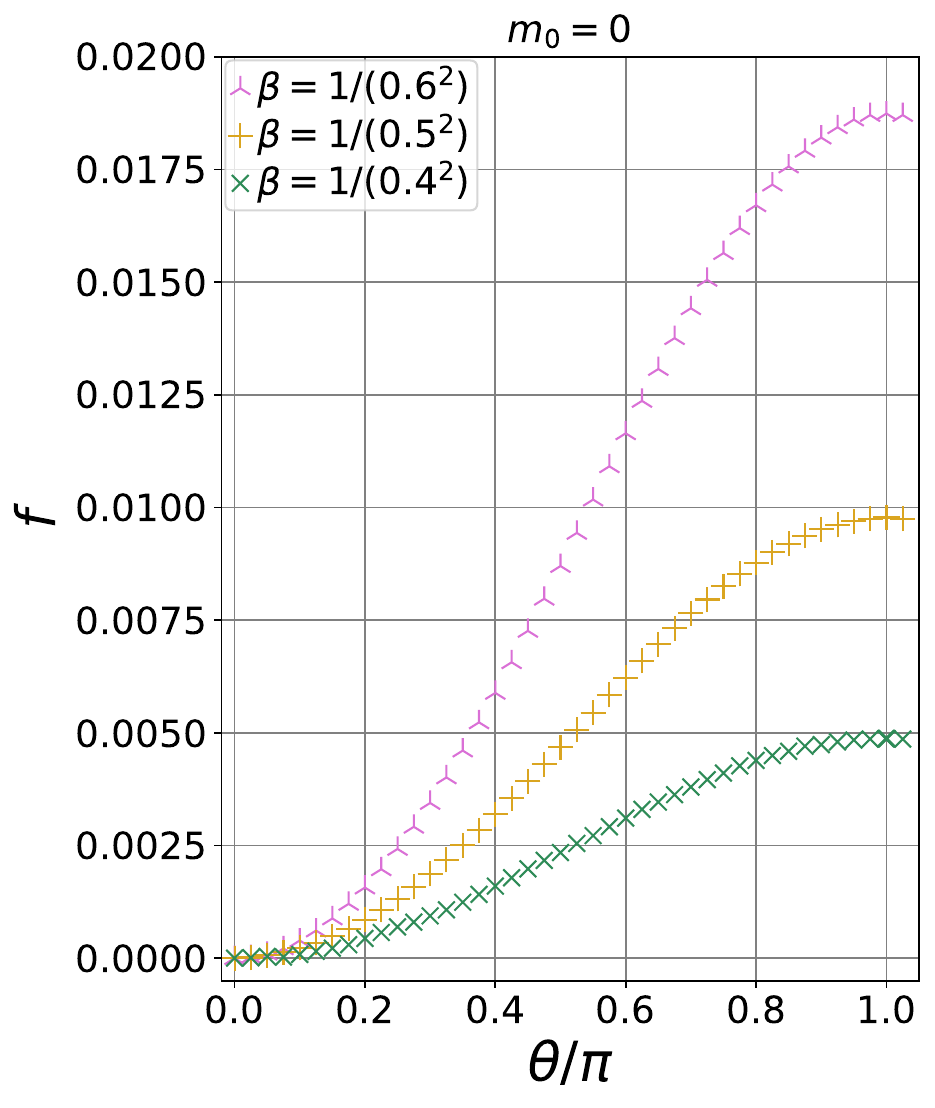}
    \caption{
         Free energy density as a function of $\theta/\pi$ at $\sqrt{\beta m_0^2}=0$ with various $\beta$. Note that the plot of $\beta=1/(0.5^2)=4$ is also depicted in Fig.~\ref{fig:mass_f}.
    }
    \label{fig:f_artifact5}
\end{figure}

From now on, we investigate the finite-mass effect.
Fig.~\ref{fig:mass_f} shows the free energy density for different masses at $\beta=1/(0.5)^2=4$.
Since we have already confirmed that our computation preserves the $2\pi$ periodicity, we just show the result in the range of $\theta\in[0,\pi]$.
With $\sqrt{\beta m^{2}_{0}}=100$, the numerical result is consistent with the large mass limit, which is described by the pure Maxwell theory.~\footnote{
    The analytic result for the lattice pure Maxwell theory is obtained by the numerical integration of
    \begin{align}
        -\beta\frac{\log Z(\theta)}{L^2} =&  -\beta\log \int_{-\pi}^{\pi}\frac{{\rm d}A_p}{2\pi}\mathrm{e}^{\beta\sum_p \cos(A_p) -\frac{
        {\rm i}\theta}{2\pi} \sum_p A_p } \ ,
        \label{eq:lat_exact_result}
    \end{align}
    which is given in Ref.~\cite{Wiese:1988qz}.
    This function corresponds to the orange line which described as ``analytic (lattice)" in Fig~\ref{fig:mass_f}.
    See Appendix~\ref{app:large} for a further explanation of the large mass limit.}
This is a validation of our numerical approach.
Decreasing the mass, we observe a clear deviation from the pure Maxwell theory due to the finite-mass effect.
This is a direct benefit of the application of the Grassmann tensor network; there is no difficulty in dealing with massive fermions in contrast to the world-line approach~\cite{Gattringer:2015nea,Goschl:2017kml} and the previous TRG approach~\cite{Butt:2019uul}.
We can also see that the free energy density tends to be smooth with respect to $\theta$ at $\theta=\pi$ by further decreasing the mass.
This situation is similar to the single-flavor Schwinger model with a $\theta$ term, where the critical endpoint appears.
However, we expect that there is no critical endpoint in the two-flavor model in the continuum limit, according to Ref.~\cite{Dempsey:2023gib}.
In addition, from the right panel of Fig.~\ref{fig:mass_f}, the result at $m_0=0$ depends on $\theta$.
This behavior does not agree with that in the continuum limit, where the free energy should be independent of $\theta$ at $m=0$ because the $\theta$ parameter can be changed to arbitrary value through the $U(1)_A$ ABJ-type anomaly.
Furthermore, our results at $\sqrt{\beta m_0^2}=0.01$ have a stark deviation from those in the mass perturbation theory with the same mass, which is shown as a curve in the right panel of Fig.~\ref{fig:mass_f}. 
These inconsistencies may be explained by the finite-$\beta$ effect.
Indeed, as seen in Fig.~\ref{fig:f_artifact5}, the free energy density at $m_0=0$ depends significantly on $\beta$, which indicates that our results with a small mass suffer from the finite-$\beta$ effect.
Also, from Fig.~\ref{fig:f_artifact5}, it is found that the $\theta$ dependence at $m_0=0$ is more enhanced with a smaller $\beta$, which could imply that the finite $\beta$ causes the strange behavior of our results with a small mass.~\footnote{
    This finite-$\beta$ effect is also pointed out by a Monte Carlo study with the staggered fermions~\cite{Azcoiti:2017mxl}.
}
To control the finite-$\beta$ effect in the small mass regime, it is necessary to simulate with at least $\beta > 6.25$ 
since there remains the $\theta$ dependence at $m_0=0$ even with $\beta=6.25$. 
Here, we note that the $O(a)$ correction 
implemented through a mass parameter shift in the Hamiltonian approach~\cite{Dempsey:2022nys} cannot be applied to our case due to the different pattern of the remnant chiral symmetry. 
The detailed discussion is presented in Appendix~\ref{app:remnant_U1A}.
Simulations for larger $\beta$ might be achieved by improving the BTRG algorithm according to Refs.~\cite{PhysRevE.97.033310,Morita:2024lwg}.
We leave this for future work.

\subsection{Topological charge density}
\label{subsec:top_charge}

\begin{figure}[htbp]
    \centering
    \includegraphics[width=0.48\hsize]{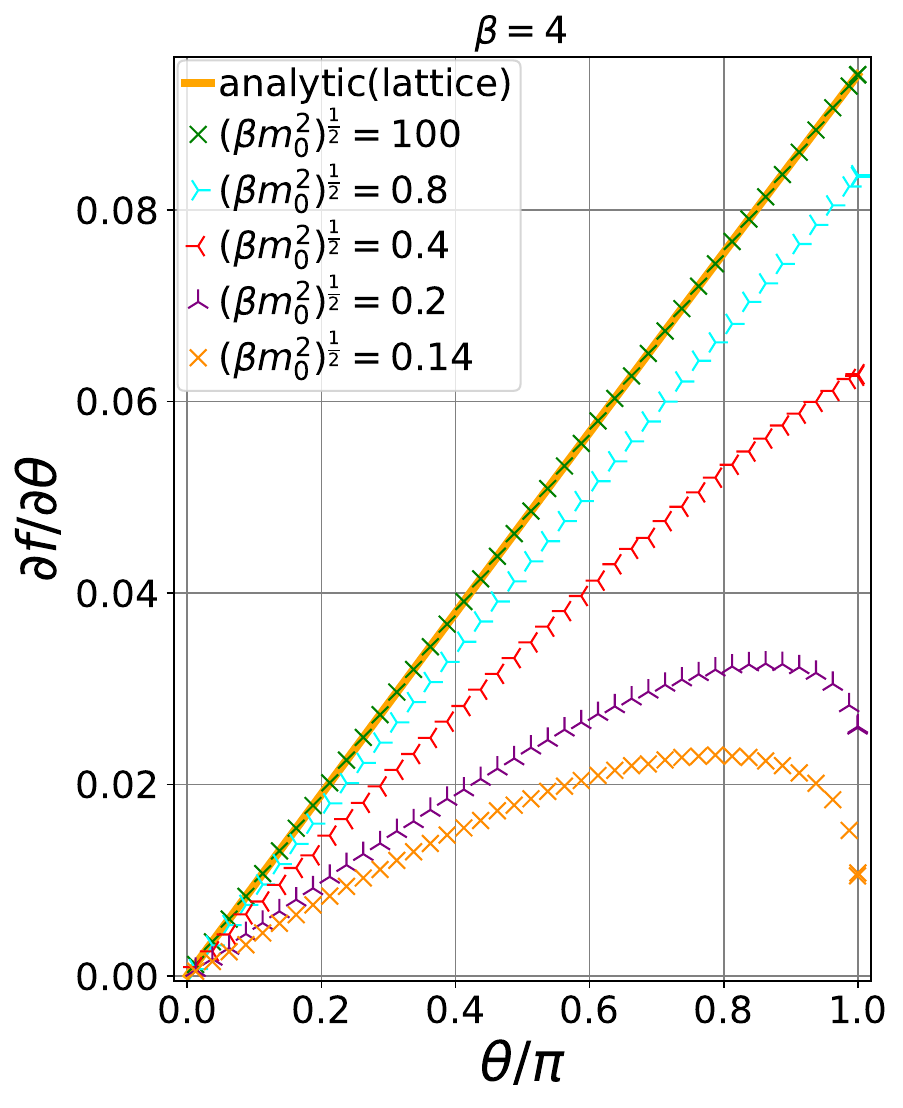}
    \includegraphics[width=0.5\hsize]{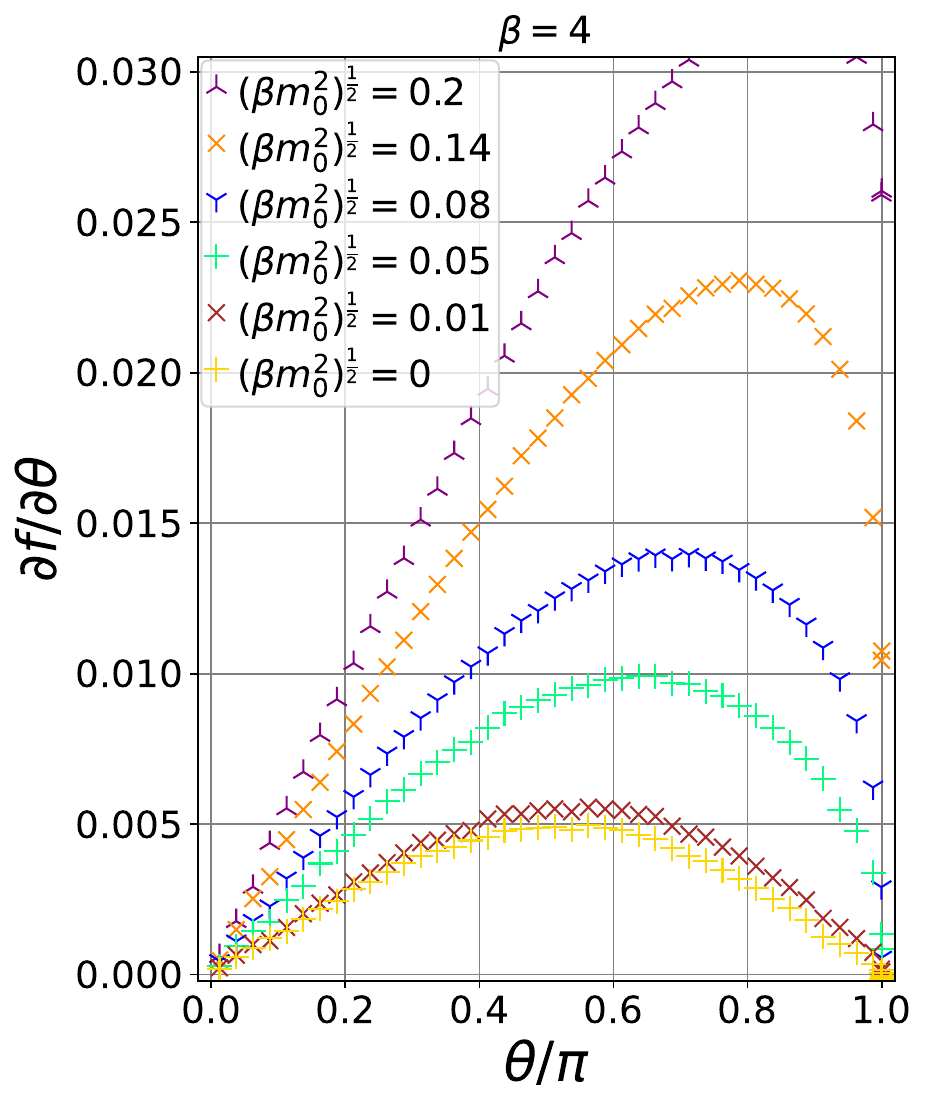}
    \caption{
         Topological charge density as a function of $\theta/\pi$ at $\beta=4$. 
    }
    \label{fig:mass_Q_all}
\end{figure}

Next, we investigate the topological charge density, which is obtained from Eq.~\eqref{eq:shifted_free_energy} by numerical differentiation in terms of $\theta$.
Fig.~\ref{fig:mass_Q_all} shows the resulting topological charge density at $\beta=4$ for different masses.
With an extremely large mass, $\sqrt{\beta m_0^2} = 100$, the topological charge depends on $\theta$ linearly, which is again consistent with the pure Maxwell theory as explained in Sec.~\ref{subsec:free_energy}.
Around $\sqrt{\beta m_0^2}\sim 0.4$, the linearity is no longer seen and the curvature appears.
From $\sqrt{\beta m_0^2}\sim 0.2$, a convex begins to appear at $\theta \approx 0.8 \pi$.
With $\sqrt{\beta m_0^2}\gtrsim0.14$, we observe discontinuities in the topological charge density, which support the first-order phase transition line at $\theta=\pi$.
Continuing to decrease $m_{0}$, the discontinuity seems to disappear.
As we argued in Sec.~\ref{subsec:free_energy}, this behavior could be due to the finite-$\beta$ effect.

\begin{figure}[htbp]
    \centering
    \includegraphics[width=0.98\hsize]{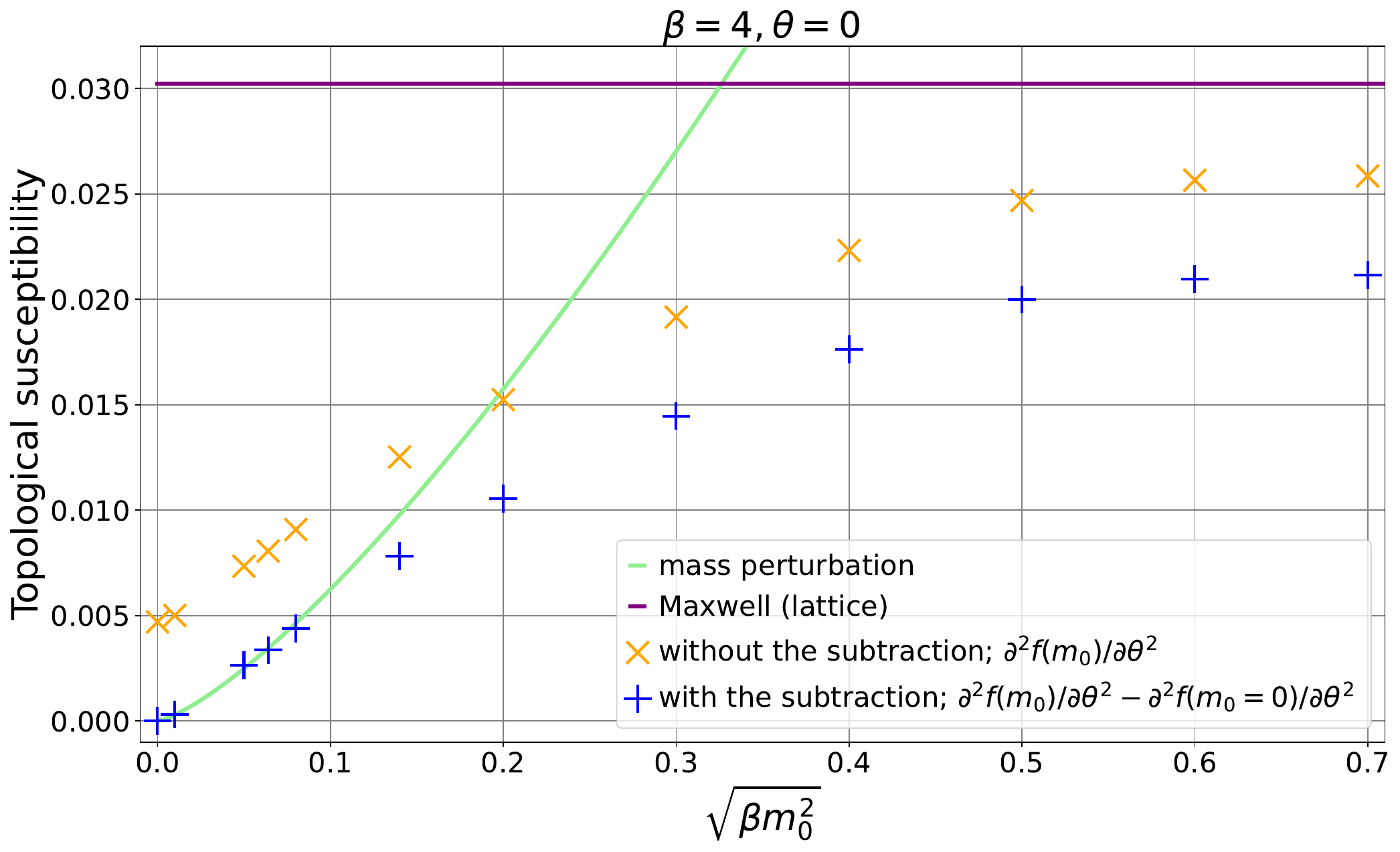}
    \caption{
        Topological susceptibility at $\theta=0$.
    }
    \label{fig:susceptibility}
\end{figure}

Here, we also show the mass dependence of the topological susceptibility at $\theta=0$ in Fig.~\ref{fig:susceptibility}.
Note that the topological susceptibility is defined by ${\partial^{2}f}/{\partial\theta^{2}}$, which can also be evaluated by the numerical differentiation.  
To stabilize the numerical differentiation, we first obtain the averaged free energy density $\bar{f}(\theta)$ as $\bar{f}(\theta)=\left\{f(\theta-\delta)+f(\theta)+f(\theta+\delta) \right\}/3$ with $\delta=0.025\pi$.
We then perform the numerical differentiation for $\bar{f}$ to evaluate the topological susceptibility at $\theta=0$ via $\{\bar{f}(\Delta)-2\bar{f}(0)+\bar{f}(-\Delta)\}/\Delta^{2}$ with $\Delta=0.075\pi$.
This prescription reduces the numerical instability and results in a smooth plot against $\sqrt{\beta m_0^2}$ as shown in Fig.~\ref{fig:susceptibility}.

Our numerical results (orange points) are consistent with the previous Monte Carlo study of the $N_{f}=2$ Schwinger model with staggered fermions at $\beta=4$~\cite{Durr:2003xs}; the numerical result shown in Fig.~\ref{fig:susceptibility} is consistent with Fig.~6 in Ref.~\cite{Durr:2003xs}.~\footnote{
    We can also see the Monte Carlo result of the topological susceptibility for the $N_{f}=2$ Schwinger model with domain-wall fermions at $\beta=1$ in Ref.~\cite{Fukaya:2004kp}.
}
We can also see that our results tend to reach the large mass limit (purple line) as we increase the mass. 
Although the results with the larger masses are not shown in this figure, it is guaranteed that the topological susceptibility approaches the limit since the free energy itself does.
Therefore, the current BTRG computation with $D=120$ seems to be sufficiently accurate to investigate the lattice model even in the vicinity of the massless limit, where the finite-$D$ effect is usually enhanced~\cite{Yoshimura:2017jpk,Akiyama:2020sfo}.
The discrepancy between the numerical results and the mass perturbation could be solely attributed to the finite-$\beta$ effect.

Furthermore, we also consider the improvement of the topological susceptibility, particularly in the small mass regime. 
In the massless case of the continuum theory, there should be no $\theta$ dependence in the free energy, thus the residual topological susceptibility at $m_0=0$ in Fig.~\ref{fig:susceptibility} is considered as a lattice artifact. 
According to the Symanzik's lattice effective theory~\cite{Symanzik:1979ph} and assuming the small mass perturbation, we can improve the result by subtracting the value of the lattice artifact at $m_0=0$. 
We also show this improved result in Fig.~\ref{fig:susceptibility} as blue plots. 
The improved ones are consistent with the mass perturbation line in small mass regime $\sqrt{\beta m_0^2}\lesssim 0.08$. 
This is another support to justify our result in the small mass regime.
Notice, however, that this improvement is based on the small mass perturbation, so it is not valid for the large mass regime.

\subsection{Ground-state degeneracy}
\label{sec:wide}

We investigate the degeneracy of the ground state at $\theta=0$ and around $\theta=\pi$.
The Schwinger model exhibits the spontaneous CP symmetry breaking, which leads to the two-fold degenerate ground state at $\theta=\pi$. 
We examine whether such a behavior also appears in the numerical calculation. 
To obtain the degeneracy, we employ the fixed-point tensor~\cite{PhysRevB.80.155131} for Grassmann variables.
Suppose we have a renormalized local Grassmann tensor $\mathcal{T}_{XT\bar{X}\bar{T}}$ using the BTRG algorithm, where $X$, $T$, $\bar{X}$, $\bar{T}$ are the Grassmann variables introduced by the coarse-graining procedure.
Using the renormalized Grassmann tensor, we define the following Grassmann matrix,
\begin{align}
\label{eq:Grassmann_TM}
    \mathcal{A}_{T\bar{T}}
    =
    \int
    {\rm d}\bar{X}{\rm d}X
    {\rm e}^{-\bar{X}X}
    \mathcal{T}_{XT\bar{X}\bar{T}}
    .
\end{align}
The ground-state degeneracy is then obtained via
\begin{align}
\label{eq:x}
    \frac{\left({\rm gTr}\mathcal{A}\right)^{2}}{{\rm gTr}\left(\mathcal{A}\right)^{2}}
    ,
\end{align}
because this quantity counts the degeneracy of the local Grassmann tensor, which corresponds with the ground-state degeneracy after the sufficient times of coarse-graining~\cite{PhysRevB.80.155131}.
Note that $\left(\mathcal{A}\right)^{2}$ in the denominator of Eq.~\eqref{eq:x} means the Grassmann matrix product.

\begin{figure}[htbp]
    \centering
    \includegraphics[width=0.48\hsize]{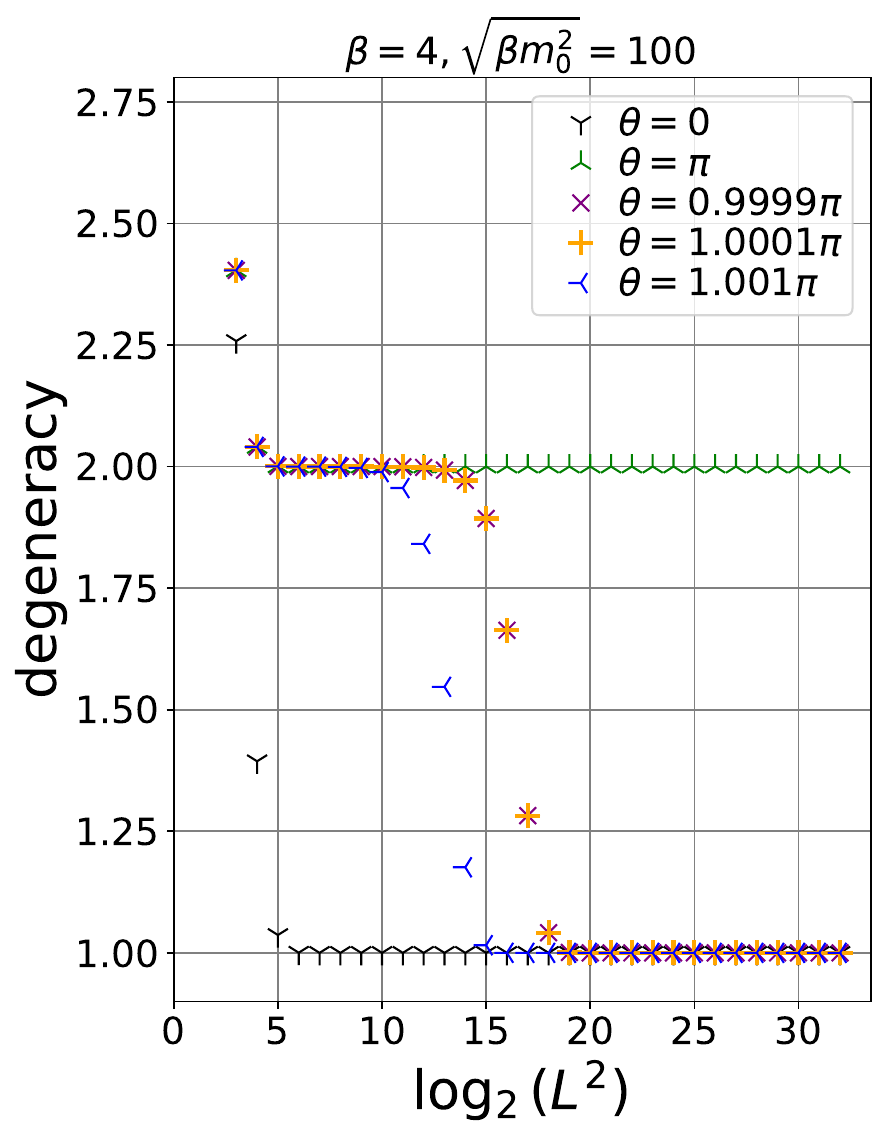}
    \includegraphics[width=0.48\hsize]{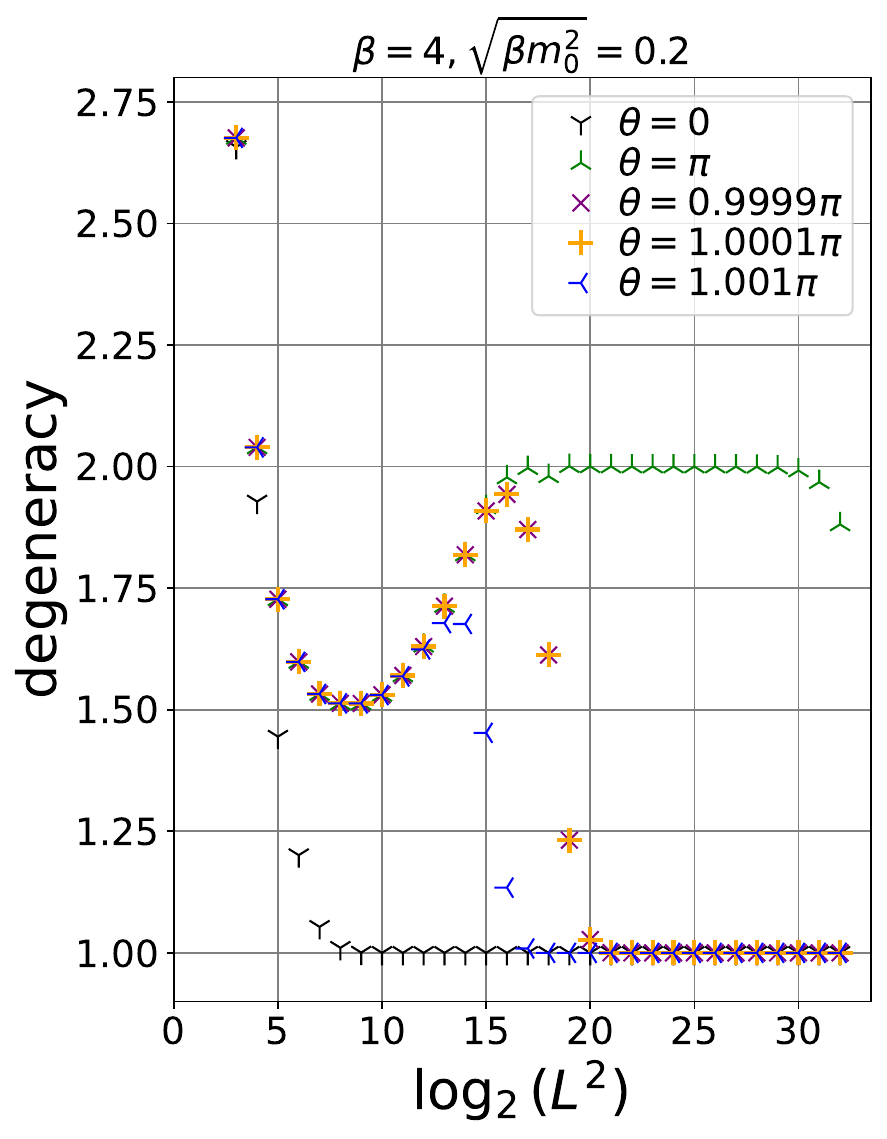}
    \caption{
        Ground state degeneracy, Eq.~\eqref{eq:x}, as a function of the coarse-graining steps at $\sqrt{\beta m_0^{2}}=100$ (left) and $\sqrt{\beta m_0^{2}}=0.2$ (right).
    }
    \label{fig:degeneracy_bm10000}
\end{figure}

Fig.~\ref{fig:degeneracy_bm10000} shows the numerical results of Eq.~\eqref{eq:x} as a function of the coarse-graining steps in the BTRG.
At $\sqrt{\beta m_0^{2}}=100$, we observe a clear plateau of $2$ at $\theta=\pi$, which indicates that the symmetry breaking takes place and the ground state is doubly degenerate at a large mass.
We also show the degeneracy at $\theta=0$, which rapidly converges to $1$ for all mass parameters. 
This is in contrast to the case around $\theta=\pi$, where the plateau does not appear unless the system size is sufficiently large.
Our result shows that the ground-state degeneracy is sensitive to $\theta$ and the system results in a unique vacuum except at $\theta=\pi$; Only a slight displacement from $\theta=\pi$ induces the unique ground state.
It should be emphasized that these results provide non-trivial evidence of the doubly degenerated ground state even at finite large mass; the two-fold degeneracy at $\theta=\pi$ is analytically rigorously proven only in the large mass limit.
We also remark that the realization of such two-fold degeneracy only at $\theta=\pi$ reflects the preserved $2\pi$ periodicity with respect to $\theta$.
However, we observe no ground-state degeneracy in the range of $\sqrt{\beta m_0^2}\le 0.14$ at $\beta=4$.
As we discussed in Sec.~\ref{subsec:free_energy}, this would be due to the finite-$\beta$ effect.
We will further investigate how the finite-$\beta$ effect modifies the phase diagram in future work.

\subsection{Correlation length}
\label{sec:mass_gap}
Finally, we investigate the correlation length at $\theta=\pi$ in the small mass region. 
According to the recent field theoretical argument~\cite{Dempsey:2023gib}, the correlation length $\xi$ is expected to scale as $\xi\sim{\rm e}^{A/(\beta m^{2}_{0})}$, where $A$ is a constant factor and analytically obtained as $A\simeq0.111$. 
Also, in the paper, this theoretical prediction is confirmed  by an MPS simulation, in which the correlation length is evaluated (up to a multiplicative factor) using the spatial volume dependence of the central charge $c$ derived from the entanglement entropy~\cite{Calabrese:2009qy}.
In this study, we estimate the correlation length within the path integral formalism using the behavior of the central charge $c$ derived from the largest eigenvalue of the transfer matrix, $\lambda_{0}$.

We find that our numerical result of $c$ lies around $c=1$ on a small volume and moves away from $c=1$ on a large volume. 
Such behavior can be explained by the finite volume effect associated with $c=1$ CFT that appears when the volume is smaller than the correlation length $\xi/a$. 
We thus define the correlation length from the volume on which $c$ begins to deviate from $1$. 
The detail derivation of the correlation length and the result of the central charge is shown in Appendix \ref{app:mass_gap}.

Fig.~\ref{fig:mass_gap} shows the logarithm of the resulting correlation length as a function of $1/(\beta m^{2}_{0})$, where we have employed slightly larger values of $\beta$ than in the previous sections to further suppress finite-$\beta$ effects.
In this figure, lines parallel to $\log\xi=A/(\beta m^{2}_{0})$ with $A\approx0.111$ are shown as gray dashed ones.
The results for $\beta\gtrsim4$ represent linear behavior alongside these lines at $1/(\beta m^{2}_{0})\gtrsim 10$, which suggest that the exponentially large correlation length with $A\approx0.111$ {is} restored in the smaller mass region for a sufficiently large $\beta$.
The obtained correlation length is also comparable with the result in Ref.~\cite{Dempsey:2023gib}, as shown in Fig.~3 of the reference. 
Such an exponentially large correlation length, equivalent to an exponentially small mass gap,  reflects its nearly conformal behavior corresponding to the $c=1$ CFT.
\begin{figure}[htbp]
    \centering
    \includegraphics[width=0.98\hsize]{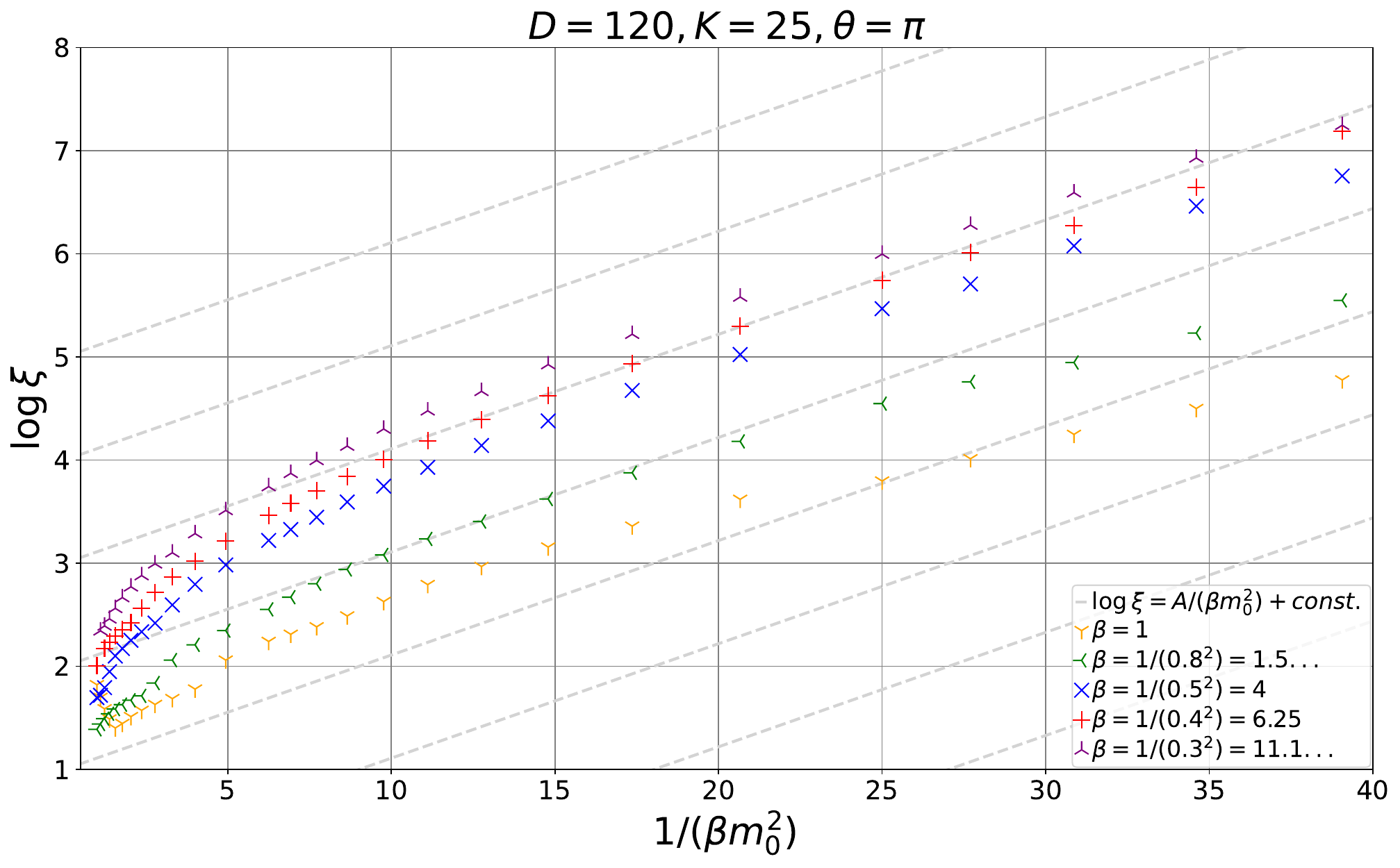}
    \caption{
        Correlation length at $\theta=\pi$ as a function of $1/(\beta m^{2}_{0})$. 
        The gray dashed lines denote lines parallel to $\log\xi=A/(\beta m^{2}_{0})$ with $A\approx0.111$.
    }
    \label{fig:mass_gap}
\end{figure}

\section{Conclusion}
\label{sec:conclusion}

We investigated the two-flavor Schwinger model with the $\theta$ term.
In our computation, we used the 2D staggered fermion action and the U(1) Wilson gauge action. 
Also, the logarithmic form was adopted for the $\theta$ term, where the $2\pi$ periodicity with respect to the $\theta$ parameter was guaranteed.
The tensor network representation was derived based on the Gauss-Legendre quadrature for the gauge field and the Grassmann tensor network formulation for the staggered fermions. 
We also employed the bond-weighted TRG algorithm to improve the accuracy of our numerical results.

We confirmed that our numerical results of the free energy density are $2\pi$ periodic for the $\theta$ parameter.
Using the large mass limit, we made a validation of our numerical approach.
Another validation was made by investigating the mass dependence of the topological susceptibility, which was in agreement with the previous Monte Carlo study in Ref.~\cite{Durr:2003xs}.
The free energy density and topological charge density were obtained in a broad range of mass.
A smooth connection of the $\theta$ dependence between the linear behavior for the large mass and the convex shape for the small mass was observed in the topological charge density.  
We found that both the free energy density and topological charge density tend to be smooth at $\theta=\pi$ by decreasing the mass.
Particularly, the results at $m_0=0$ show a non-negligible $\theta$ dependence which should be absent in the continuum limit.
We also checked that the two-vacua degeneracy is realized at $\theta=\pi$, as a consequence of the $2\pi$ periodicity, in the large mass regime.
The degeneracy cannot be found in the smaller mass regime.
These observations strongly suggest that the phase diagram at finite $\beta$ be different from that in the continuum limit.
This possibility will be addressed elsewhere.
It would also be interesting to investigate the same model but with the $N_{f}=2$ Wilson fermions as another future work, because the staggered fermion involves the so-called taste-breaking effect, where the flavor and chiral symmetries are violated by the discretization effect.
Finally, we estimated the correlation length from the volume dependence of the central charge, of which the result suggests that it tends to be exponentially large in the smaller mass region.
We also confirmed that the resulting correlation length with $\beta\gtrsim4$ is consistent with the recent theoretical argument and the MPS simulation~\cite{Dempsey:2023gib}.

We expect that our results will get closer to the continuum limit with the larger $\beta$. 
To approach the continuum limit, it will be necessary to enlarge both the bond dimension $D$ and the cutoff parameter $K$.
We are planning to combine our Grassmann BTRG algorithm with the randomized SVD~\cite{PhysRevE.97.033310,Morita:2024lwg}, which reduces both the computational memory and time.
This improved algorithm will bring us one step closer to the more precise study of the finite-$\beta$ effect, the smaller mass region, and the continuum limit.
We are also planning to extend our study by employing the L\"{u}scher gauge action, which improves the convergence to the continuum limit with respect to $\beta$ compared with the standard Wilson gauge action. 
Although the L\"{u}scher gauge action~\cite{Luscher:1998du} is known to cause the so-called topology freezing in the Monte Carlo simulation, some recent studies have numerically confirmed that such an issue is automatically resolved in the TRG computations~\cite{Akiyama:2024qer,Akiyama:2025yxh}.
It seems intriguing to explore whether the mass dependence of the free energy in Eq.~\eqref{eq:f_exact_light_result} could be altered in the small mass region by the exponentially small mass gap we observed.
These studies will be reported elsewhere.

\section*{Acknowledgement}

The numerical results presented in this paper were completed by the computing system of Yukawa Institute for Theoretical Physics, Kyoto University (Sushiki server and Yukawa-21).
HK would like to thank Shigeki Sugimoto and Yuya Tanizaki for their useful discussions.
This work started during the workshop held at Yukawa Institute for Theoretical Physics (YITP-W-22-13).
The work of HK was supported by the establishment of university fellowships towards the creation of science technology innovation and RIKEN Special Postdoctoral Researcher Program.
SA acknowledges the support from JSPS KAKENHI Grant Number JP23K13096, the Endowed Project for Quantum Software Research and Education, the University of Tokyo~\cite{qsw}, the Center of Innovations for Sustainable Quantum AI (JST Grant Number JPMJPF2221), and the computational resources of Wisteria/BDEC-01 and Cygnus and Pegasus under the Multidisciplinary Cooperative Research Program of Center for Computational Sciences, University of Tsukuba. 
KM is supported in part by Grants-in-Aid for JSPS Fellows (Nos.\ JP22J14889, JP22KJ1870) and by JSPS KAKENHI with Grant No.\ 22H04917.
ST is supported in part by JSPS KAKENHI Grants No. 21K03531, and No. 22H01222.
This work was supported by MEXT KAKENHI Grant-in-Aid for Transformative Research Areas A “Extreme Universe” No. 22H05251.

\appendix

\section{Analytic calculation of $N_f=2$ Schwinger model}
\label{app:analytic}

Physical quantities of the Schwinger model can be exactly calculated in the large mass limit and the massless case.
The large mass limit just corresponds to the Maxwell theory, which is exactly solvable. The massless case can be analyzed by the Abelian bosonized action~\cite{Coleman:1976uz}, furthermore, we can include a small mass term for it perturbatively.
In this appendix, we analytically calculate the free energy.

\subsection{Large mass limit}
\label{app:large}

In the $m\to\infty$ limit, all fermions are decoupled in the IR. 
This theory goes to the 2D pure $U(1)$ gauge theory, which is called the 2D Maxwell theory.
In two dimensions, the Maxwell theory does not have any propagating degrees of freedom, so we can solve it by hand.

The effective action in the $m\to\infty$ limit is,
\begin{align}
  S_M =& \int {\rm d}^2x \, \bigg\{\frac{1}{4g^2}F_{\mu\nu}F^{\mu\nu} + \frac{{\rm i}\theta}{4\pi}\epsilon^{\mu\nu}F_{\mu\nu}  \bigg\} \ .
\end{align}
Taking the $A_1=0$ gauge, $F_{12}=-F_{21}=\partial_1A_2$, the action only includes $F_{12}$ as
\begin{align}
  S_M =& \int {\rm d}^2x \, \bigg\{\frac{1}{2g^2}F_{12}^2 + \frac{{\rm i}\theta}{2\pi} F_{12}  \bigg\} \ .
\label{MaxwellS}
\end{align}
It is easy to integrate out $F_{12}$, and we can calculate the partition function. 
Let us consider the theory on a torus $T^2$. 
Since we need to treat the instanton sector carefully, we describe $A_{2}$ separately as
\begin{align}
  A_2 =& \frac{2\pi n}{V}x_1 + w(x_1,x_2) \ ,& (n\in \mathbb{Z})
\label{gauge}
\end{align}
where $w(x_1, x_2)$ is a periodic function for both direction $x_1$ and $x_2$. 
Therefore, $w(x_1+L_1, x_2)=w(x_1, x_2+L_2)=w(x_1, x_2)$, where $L_1$ and $L_2$ are the system length of each directions and $V=L_1L_2$. 
The first term of Eq.~(\ref{gauge}) is not periodic for $x_1$, but this is well-defined. 
We require the well-definedness of $\int {\rm d}^2x \, F_{12}$ up to $2\pi\mathbb{Z}$ because the partition function includes $\exp{\left(\frac{{\rm i}\theta}{2\pi}\int {\rm d}^2x \, F_{12}\right)}$ and it cannot distinguish the $2\pi\mathbb{Z}$ difference of $\int {\rm d}^2x \, F_{12}$. 
Therefore, $\int {\rm d}^2x \, F_{12}$ with Eq.~(\ref{gauge}) has the $2\pi\mathbb{Z}$ ambiguity for its boundary condition, and we write this part as a term proportional to $x_1$. The integer $n$ in this term is nothing but the origin of the instanton number.
We can evaluate the instanton number for Eq.~(\ref{gauge}) as,
\begin{align}
    \frac{1}{2\pi}\int {\rm d}^2x\, F_{12} =& \frac{1}{2\pi}\int {\rm d}^2x\, \partial_{1}\left(\frac{2\pi n}{V}x_1 + w(x_1,x_2) \right) \nn \\
    =&  n + \frac{1}{2\pi}\int {\rm d}x_2\, \left[w(x_1=L_1,x_2)-w(x_1=0,x_2) \right] = n \ ,
  \label{instanton}
\end{align}
and $w(x_1, x_2)$ does not affect the instanton number.~\footnote{
    This $w$ corresponds to the $\mathbb{R}$-valued gauge field in non-compact QED.
}
To consider the path integral for Eq.~(\ref{MaxwellS}) with the ansatz (\ref{gauge}), we should treat the instanton number $n$ carefully. The partition function of Eq.~(\ref{MaxwellS}) becomes,
\begin{align}
  Z(\theta) =& \int \mathcal{D}A \, \mathrm{e}^{-S_M} \nn\\
  =& \sum_n \int \mathcal{D}w \, \mathrm{e}^{ \left\{-\int {\rm d}^2x\,\frac{1}{2g^2}\left(\partial_1 w\right)^2 -\frac{2\pi^2}{g^2V}n^2 -{\rm i}\theta n \right\}} \nn\\
  =& C' \sum_n \mathrm{e}^{ -\frac{2\pi^2}{g^2V}n^2 -{\rm i}\theta n } \nn\\
  =& C'' \sum_n \mathrm{e}^{ -\frac{g^2V}{8\pi^2}\left(\theta -2\pi n \right)^2 } \ .
  \label{theta_func}
\end{align}
The overall factor of the partition function is not important for our purpose and we just write it as $C'$ and $C''$.
We use the Poisson summation formula to rewrite the infinite summation. We can evaluate this partition function (\ref{theta_func}) in the large volume limit, where the Gaussian factor in Eq.~(\ref{theta_func}) is highly suppressed, and the contribution of the smallest $(\theta -2\pi n)^2$ term is dominant in the summation. 
Although the partition function depends on the volume $V$, the free energy density does not depend on $V$. 
Therefore, we evaluate the free energy density as,
\begin{align}
    -\frac{\log Z(\theta)}{V} =& -\frac{\log{C''}}{V}-\frac{1}{V}\log \sum_n \mathrm{e}^{ -\frac{g^2V}{8\pi^2}\left(\theta -2\pi n \right)^2 } \nn \\
    \simeq& -\frac{\log{C''}}{V} + \min_n  \frac{g^2}{8\pi^2}\left(\theta -2\pi n \right)^2 \ .
    \label{f_Maxwell}
\end{align}
This evaluation is exact in the $V\to\infty$ limit. 
The first term of Eq.~(\ref{f_Maxwell}) is just a constant, so it is irrelevant to our study. 
The second term describes the $\theta$ dependence of the free energy density. 

\begin{figure}[htbp]
    \centering
    \includegraphics[width=0.49\hsize]{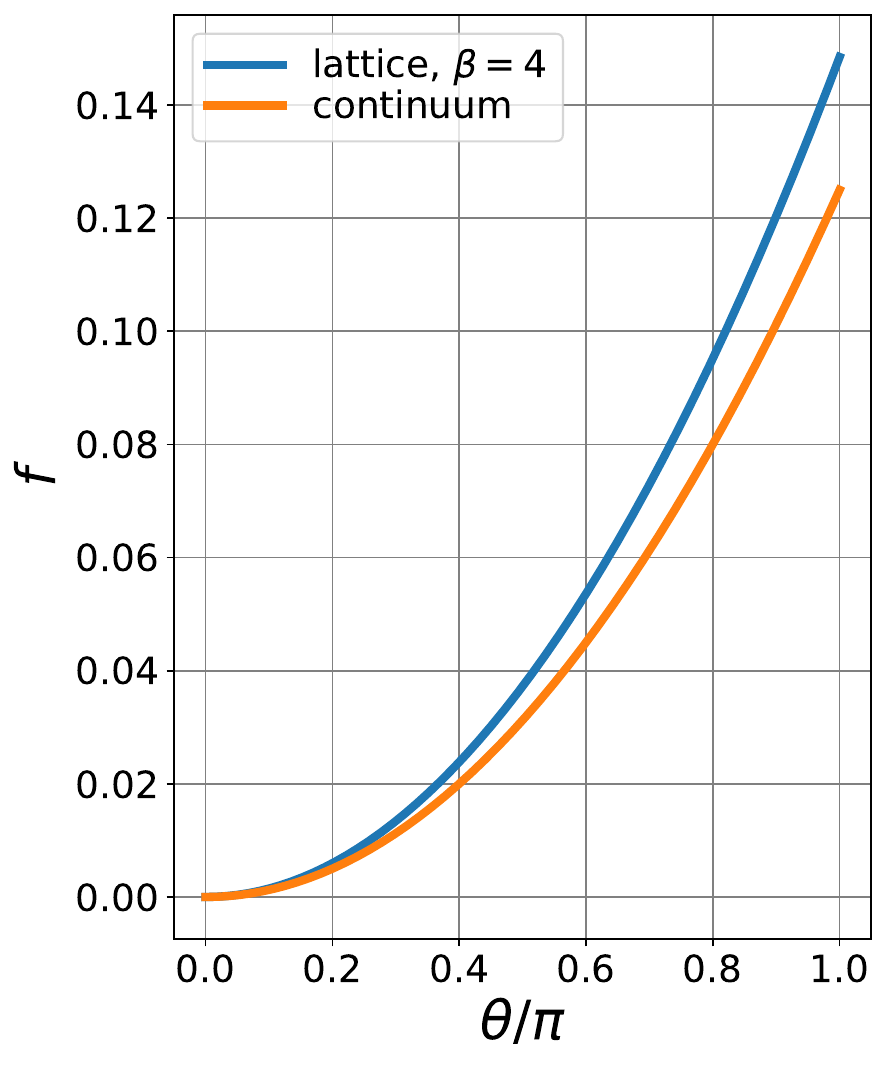}
    \caption{
        The analytic result of the 2D Maxwell theory.
        The $\theta$ dependence of the free energy density for the continuum and lattice actions are plotted. 
    }
    \label{fig:Maxwell}
\end{figure}

We can also evaluate the free energy on a lattice.
The lattice action for the 2D Maxwell theory can be written as
\begin{align}
	S_{\rm lat}=&
	-\beta\sum_{n\in\Lambda_{2}}
	\Re\left[U_{1}(n)U_{2}(n+\hat{1})U^{*}_{1}(n+\hat{2})U^{*}_{2}(n)\right] \nn\\
	&-
    \frac{\theta}{2\pi}\sum_{n}
	\log\left[U_{1}(n)U_{2}(n+\hat{1})U^{*}_{1}(n+\hat{2})U^{*}_{2}(n)\right] \nn\\
    =& -\beta\sum_p \cos(A_p) -\frac{{\rm i}\theta}{2\pi} \sum_p A_p \ ,
\end{align}
as in section~\ref{sec:lattice}.
We can evaluate the free energy density for this action analytically, as written in Refs.~\cite{Wiese:1988qz, Kuramashi:2019cgs}:
\begin{align}
	Z_{\rm lat}=&
	\sum_Q \left(z_p(\theta+2\pi Q, \beta) \right)^V \ , \\
	z_p(\theta, \beta) =& \int_{-\pi}^{\pi}\frac{{\rm d}A_p}{2\pi}\mathrm{e}^{\beta \cos(A_p) +\frac{{\rm i}\theta}{2\pi} A_p } \ .
    \label{lat_exact}
\end{align}
Calculating Eq.~(\ref{lat_exact}) numerically, we can make a comparison with the TRG results.
Note that the log of Eq.~(\ref{lat_exact}) is slightly different from the continuum result in Eq.~(\ref{f_Maxwell}). 
We show the plot at $\beta=4$ in Fig.~\ref{fig:Maxwell}, as an example.
In the continuum limit ($\beta\to\infty$), these two should have the same value.

\subsection{Small mass regime}
\label{app:small}

This appendix aims to calculate the free energy density of this action (\ref{contS}). 
It is difficult to analyze the fermionic action directly, so we change this action to the abelian bosonized action.

To consider abelian bosonization, we need the bosonization rules, which are known as,
\begin{align}
    \label{bosonization_rule}
    \bar{\psi}\gamma^{\mu}\psi \, \longleftrightarrow& ~\, -\frac{\mathrm{i}}{2\pi}\epsilon^{\mu\nu}\partial_{\nu}\phi \nn \\
    \bar{\psi}{\rm i}\gamma^{\mu}\partial_{\mu}\psi \,\longleftrightarrow& ~\, \frac{1}{8\pi}\partial^{\mu}\phi\partial_{\mu}\phi  \\
    \bar{\psi}\psi \, \longleftrightarrow& ~\, Cm'\mathcal{N}_{m'} \cos(\phi) \ , \nn
\end{align}
according to Ref.~\cite{Coleman:1974bu}.
In the relation.~\eqref{bosonization_rule}, $\mathcal{N}_{m'}$ is the normal ordering for the scale $m'$. 
Note that $m'$ is just a parameter and we can choose an arbitrary value.
Using this dictionary, the action (\ref{contS}) can be written as
\begin{align}
    S =& \int {\rm d}^2x \bigg\{\frac{1}{4g^2}F_{\mu\nu}F^{\mu\nu} + \frac{{\rm i}}{4\pi}\left(\phi_1+\phi_2+\theta \right)\epsilon^{\mu\nu}F_{\mu\nu} +  \frac{1}{8\pi}\big(\partial_{\mu}\phi_1\partial^{\mu}\phi_1+\partial_{\mu}\phi_2\partial^{\mu}\phi_2 \big) \nn\\
    &+ Cmm'\mathcal{N}_{m'}\big[ \cos(\phi_1)+\cos(\phi_2) \big]   \bigg\} \ ,
    \label{abelianS}
\end{align}
where
\begin{align}
\phi_1, \phi_2 \in& [0,2\pi) \ .
\label{phi2pi}
\end{align}
The numerical constant $C$ is known as $C=\mathrm{e}^{\gamma}/2\pi$~\cite{Smilga:1992hx}.~\footnote{
    The numerical value of the Euler constant, $\gamma = 0.57721\cdots$, is important for the analysis in the section~\ref{sec:results}.
}
We set our scalar fields to take their values on $[0,2\pi)$. This is a different convention from Coleman's paper~\cite{Coleman:1976uz}.~\footnote{
    This difference comes from $2\pi$ periodicity of scalar field $\phi_i$. This is the reason why the coefficient of kinetic terms of scalar fields in Eq.~(\ref{abelianS}) are not $1/2$ but $1/8\pi$ for $2\pi$ periodic scalar in Eq.~(\ref{phi2pi}). We can read the radius of compact scalars $R$ from this coefficient, as $1/8\pi = 1/2R^2$.
}

Under the change of the normal ordering scale $m$, the coefficient of the cosine-shape mass term changes as
\begin{align}
\mathcal{N}_{m} \cos(\phi) = \left(\frac{m'}{m}\right)^{\frac{R^2}{4\pi}}\mathcal{N}_{m'} \cos(\phi) \ ,
\label{NormalOrder}
\end{align}
where $R$ is a radius of the compact scalar $\phi$, and $R=2\sqrt{\pi}$ for free fermions~\cite{Coleman:1974bu}. 
Therefore, Eq.~(\ref{NormalOrder}) for the action (\ref{abelianS}) can be written as
\begin{align}
  m\mathcal{N}_{m} \cos(\phi_i) = m'\mathcal{N}_{m'} \cos(\phi_i) \ ,
\label{NO}
\end{align}
where $i$ denotes the flavor, $i=1,2$.

In the small mass regime, we consider the $m\ll g$ case. In this regime, we can integrate out the gauge field $A_{\mu}$ and one heavy scalar field. 
This heavy scalar corresponds to the $\eta'$ meson in QCD, whose mass comes from the $U(1)_A$ anomaly. 
This system is analyzed by Coleman~\cite{Coleman:1976uz}. 
In this section, we follow the calculation in Ref.~\cite{Coleman:1976uz}.

To obtain an effective theory with only one scalar field, we redefine the scalar fields as
\begin{align}
  \phi_+ =& ~\frac{1}{2}\left( \phi_1+\phi_2 + \theta \right)  \ , \nn\\
  \phi_- =& ~\frac{1}{2}\left( \phi_1-\phi_2 \right) \ . \nn
\end{align}
This redefinition does not change the $2\pi$ periodicity of the scalars. Therefore, $\phi_+, \phi_- \in [0,2\pi)$. However, it changes the radius of these compact scalars, which appears in Eq.~(\ref{NormalOrder}) as $R$, from $R=2\sqrt{\pi}$ to $R=\sqrt{2\pi}$. It is clear that the coefficient of the kinetic term ($=1/2R^2$) changes from $1/8\pi$ to $1/4\pi$.

Now, the redefined field $\phi_+$ corresponds to the heavy scalar field which is decoupled in IR limit, and $\phi_-$ is a light scalar that appears in the effective action. The action, Eq.~(\ref{abelianS}), can be written by $\phi_+$ and $\phi_-$ as
\begin{align}
  S =& \int {\rm d}^2x \bigg\{\frac{1}{2g^2}F_{12}^2 + \frac{{\rm i}}{2\pi}2\phi_+F_{12} +  \frac{1}{4\pi}\big(\partial_{\mu}\phi_+\partial^{\mu}\phi_++\partial_{\mu}\phi_-\partial^{\mu}\phi_- \big) \nn\\
&+ 2Cmm'\mathcal{N}_{m'}\big[ \cos(\phi_+-\frac{\theta}{2})\cos(\phi_-) \big]   \bigg\} \ .
\end{align}
We can integrate out the gauge field $A_{\mu}$, in the same way in Eq.~(\ref{theta_func}). Here, $\phi_+$ corresponds to the $\theta$ parameter in Eq.~(\ref{theta_func}). After integrate out $A_{\mu}$, the effective action becomes,
\begin{align}
    S =& \min_n \int {\rm d}^2x \bigg\{ \frac{1}{4\pi}\big(\partial_{\mu}\phi_+\partial^{\mu}\phi_++\partial_{\mu}\phi_-\partial^{\mu}\phi_- \big) + \frac{g^2}{8\pi^2}(2\phi_+ -2\pi n)^2 \nn\\
    &+ 2Cmm'\mathcal{N}_{m'}\big[ \cos(\phi_+-\frac{\theta}{2})\cos(\phi_-) \big]   \bigg\} \nn \\
    =& \min_n \int {\rm d}^2x \bigg\{ \frac{1}{4\pi}\big(\partial_{\mu}\phi_+\partial^{\mu}\phi_++\partial_{\mu}\phi_-\partial^{\mu}\phi_- \big) + \frac{1}{4\pi}\mu^2(\phi_+ -\pi n)^2 \nn\\
    &+ 2Cmm'\mathcal{N}_{m'}\big[ \cos(\phi_+-\frac{\theta}{2})\cos(\phi_-) \big]   \bigg\} \ .
\label{mesonS}
\end{align}
where $\mu^2=2g^2/\pi$ and $\mu$ is a mass of $\phi_+$, in analogy with the mass of $\eta'$ meson comes from the Witten-Veneziano formula. 
Eq.~(\ref{mesonS}) includes only two bosons, $\phi_{\pm}$, and these masses are non-degenerate.

This system can be analyzed perturbatively for $m$ in the $m/g\ll1$ case.
We focus on the light scalar $\phi_-$, by integrating out the heavy scalar $\phi_+$, which can appear in the Feynman diagram of $\phi_-$ through the loop. 
To neglect the loop contribution of $\phi_+$, we take the normal ordering at the scale $\mu=g\sqrt{2/\pi}$. 
By the relation (\ref{NO}), we set the normal ordering at the scale of $\mu$ and integrate out $\phi_+$. Then, the action (\ref{mesonS}) becomes,~\footnote{
    Before integrating out $\phi_+$, we take normal order at the scale $m'=\mu$. 
    Since this normal ordering is for $\phi_+$ and $\phi_-$, we use Eq.~(\ref{NO}). 
    It is good to see Eq.~(\ref{abelianS}) and read the radius of the compact bosons as $R=2\sqrt{\pi}$.
    The vacuum expectation value of $\phi_+$ should be set as $\phi_+=\pi n$, to minimize the second term of Eq.~(\ref{mesonS}).
}
\begin{align}
    S =& \min_n \int {\rm d}^2x \bigg\{ \frac{1}{4\pi}\big(\partial_{\mu}\phi_-\partial^{\mu}\phi_- \big) + 2Cm\mu\cos\left(\frac{\theta-2\pi n}{2}\right)\mathcal{N}_{\mu}\big[ \cos(\phi_-) \big]   \bigg\} \ .
\label{phi_-S}
\end{align}
To obtain mass gap $m_{\Delta}$, we take normal order again for this effective action (\ref{phi_-S}) at the scale $m_{\Delta}$. 
After taking the normal order (\ref{phi_-S}), the action should be the following shape;
\begin{align}
    S =& \int {\rm d}^2x \bigg\{ \frac{1}{4\pi}\big(\partial_{\mu}\phi_-\partial^{\mu}\phi_- \big) + m_{\Delta}^2\mathcal{N}_{m_{\Delta}}\big[\cos(\phi_-) \big]   \bigg\} \ .
\label{mgapS}
\end{align}
To identify Eqs.~(\ref{phi_-S}) and (\ref{mgapS}), we solve the following equation.~\footnote{
    Here, the radius of the compact boson is changed from Eq.~(\ref{NO}), because we already integrating out $\phi_+$. The radius of $\phi_-$ is $R=\sqrt{2\pi}$, for the relation (\ref{NormalOrder}).
}
\begin{align}
    2Cm\mu\cos\left(\frac{\theta-2\pi n}{2}\right)\mathcal{N}_{\mu}\big[ \cos(\phi_-) \big] 
    =& 2Cm\mu\cos\left(\frac{\theta-2\pi n}{2}\right)\left( \frac{m_{\Delta}}{\mu} \right)^{\frac{1}{2}}\mathcal{N}_{m_{\Delta}}\big[ \cos(\phi_-) \big] \nn\\
    =& m_{\Delta}^2\mathcal{N}_{m_{\Delta}}\big[\cos(\phi_-) \big]  \ ,
\label{mgap_eq}
\end{align}
whose solution is,
\begin{align}
    m_{\Delta}
    =& \left(2Cm\sqrt{\mu}\cos\left(\frac{\theta-2\pi n}{2}\right) \right)^{\frac{2}{3}} \ .
\label{mgap}
\end{align}
Finally, the $\theta$ dependence of the free energy density is obtained as~\footnote{
    The free energy is the value of the effective action in the IR limit, where we can neglect the kinetic terms, so just the mass term remains.
}
\begin{align}
    -\frac{\log Z(\theta)}{V}
    =& \min_n m_{\Delta}^2 \nn\\
    =& \min_n \left(2Cm\sqrt{\mu}\cos\left(\frac{\theta-2\pi n}{2}\right) \right)^{\frac{4}{3}} \nn \\
    =& \min_n \left\{\left(\mathrm{e}^{\gamma}\right)^{\frac{4}{3}}\pi^{-\frac{5}{3}}2^{\frac{1}{3}}\left(\frac{m^2}{g^2}\right)^{\frac{2}{3}}g^2\cos^{\frac{4}{3}}\left(\frac{\theta-2\pi n}{2}\right) \right\} \ .
    \label{f_exact_light}
\end{align}


\section{Discrete remnant of the $U(1)_A$ symmetry on the lattice}
\label{app:remnant_U1A}

In the Hamiltonian formalism, it is known that $O(a)$ correction for staggered fermions can be absorbed into a shift of the mass term owing to the remnant chiral symmetry~\cite{Dempsey:2022nys}.
Although the exact chiral symmetry is broken in this action, a flavor-dependent $U(1)$ symmetry and certain discrete chiral symmetries remain~\cite{Golterman:1984cy,Kilcup:1986dg}. 
In this appendix, we discuss the chiral symmetry in the single-component staggered fermion in the 2D spacetime, both in the Lagrangian and Hamiltonian formalisms.

In the continuum theory, the $\theta$ parameter is shifted under a $U(1)_A$ transformation with parameter $\alpha$:
\begin{align}
	\psi &\to \mathrm{e}^{\mathrm{i}\gamma_3\alpha}\psi \ , &
    \bar{\psi} &\to \bar{\psi}\mathrm{e}^{\mathrm{i}\gamma_3\alpha} \ , &
    \theta &\to \theta + 2\alpha \ .
\label{eq:U(1)A}
\end{align}
For the massless theory, the action possesses classical $U(1)_A$ symmetry.
Since this $U(1)_A$ is anomalous, the $\theta$ parameter is also transformed as $\theta \to \theta + 2\alpha$ under the transformation and it has no physical significance.

In the Lagrangian formalism, one considers the staggered fermion on a discrete spacetime. 
Such a lattice fermion reproduces a two-flavor Dirac fermion in the continuum limit, since it contains three doublers in addition to the original fermion mode~\cite{Kogut:1974ag}.  
In this theory, the $U(1)_A$ symmetry is broken and reduced to the invariance under the transformation
\begin{align}
	\chi(n) &\to \mathrm{e}^{\mathrm{i}\epsilon(n)\beta}\chi(n) \ , &
    \bar{\chi}(n) &\to \bar{\chi}(n)\mathrm{e}^{\mathrm{i}\epsilon(n)\beta} \ , &
    \epsilon(n) = (-1)^{n_1+n_2}, \ 
\label{eq:U(1)epsilon}
\end{align} 
which we refer to as the $U(1)_{\epsilon}$ symmetry. 
This symmetry is not identical to the $U(1)_A$ symmetry in the continuum limit; the $U(1)_{\epsilon}$ transformation contains not only the flavor-independent component but also the flavor-dependent one as
\begin{align}
	\psi &\to \mathrm{e}^{\mathrm{i}\gamma_3\sigma_{\epsilon}\beta}\psi \ , &
    \bar{\psi} &\to \bar{\psi}\mathrm{e}^{\mathrm{i}\gamma_3\sigma_{\epsilon}\beta} \  ,
\label{eq:U(1)epsilonConti}
\end{align}
where $\sigma_{\epsilon}$ is a certain Pauli matrix in the $SU(2)$ flavor space.\footnote{
    $\sigma_{\epsilon}$ is defined as the same Pauli matrix as $-i\gamma_1\gamma_2$, of which the explicit form depends on the representation of the $\gamma$ matrices.     
    For the chiral representation, where $\gamma_1=\sigma_1$ and $\gamma_2=\sigma_2$, $\sigma_{\epsilon}$ is equal to $\sigma_3$. 
}
Note that this transformation does not shift the $\theta$ parameter, since two Dirac fermions with different flavors acquire opposite phases and cancel each other.\footnote{
    In other words, this transformation (\ref{eq:U(1)epsilonConti}) is not anomalous. We cannot see $U(1)_A$ anomaly from $U(1)_{\epsilon}$ transformation.
}

In the Hamiltonian formalism, only the spatial direction is discretized.
Here, a single-component staggered fermion corresponds to a one-flavor Dirac fermion in the continuum, as there is only one doubler.
The remaining classical symmetry in this action is the $\mathbb{Z}_2$ subgroup of the $U(1)_A$ symmetry.
This $\mathbb{Z}_2$ transformation shifts the $\theta$ parameter to $\theta+\pi$ through its anomaly. This effect indicates that the theory for $\theta$ is equivalent to that for $\theta+\pi$ only at the massless point, where the $O(a)$ correction can be absorbed into a shift of the mass parameter~\cite{Dempsey:2022nys}.

However, such a modification is not applicable in our setup, because we employ the Lagrangian formalism. 
As discussed above, when both spacetime directions are discretized, the single-component staggered fermion does not retain a flavor-independent $\mathbb{Z}_2$ symmetry.
Therefore, no analogous $O(a)$ mass shift exists in our case.


\section{Derivation of the local Grassmann tensor}
\label{app:derivation_tn}

We demonstrate how to derive the Grassmann tensor in Eq.~\eqref{eq:fermion_tensor}.
Firstly, we decompose the hopping terms in Eq.~\eqref{eq:lat_path_int} by introducing the auxiliary Grassmann fields as
\begin{align}
	&\exp\left[
	-\frac{\eta_{\nu}(n)}{2}\bar{\chi}(n){\rm e}^{{\rm i}\pi a_{\nu}(n)}\chi(n+\hat{\nu})
	\right]
	\nonumber\\
	&=
        \int_{\bar{\zeta}_{\nu}(n),\zeta_{\nu}(n)}
	\exp\left[
	\frac{\eta_{\nu}(n){\rm e}^{{\rm i}\pi a_{\nu}(n)}}{\sqrt{2}}\bar{\chi}(n)\zeta_{\nu}(n)
	\right]
	\exp\left[
	\frac{1}{\sqrt{2}}\chi(n+\hat{\nu})\bar{\zeta}_{\nu}(n)
	\right]
	,
\end{align}
\begin{align}
	&\exp\left[
	\frac{\eta_{\nu}(n)}{2}\bar{\chi}(n+\hat{\nu}){\rm e}^{-{\rm i}\pi a_{\nu}(n)}\chi(n)
	\right]
	\nonumber\\
	&=
        \int_{\bar{\xi}_{\nu}(n),\xi_{\nu}(n)}
	\exp\left[
	\frac{1}{\sqrt{2}}\bar{\chi}(n+\hat{\nu})\bar{\xi}_{\nu}(n)
	\right]
	\exp\left[
	\frac{\eta_{\nu}(n){\rm e}^{-{\rm i}\pi a_{\nu}(n)}}{\sqrt{2}}\chi(n)\xi_{\nu}(n)
	\right]
	,
\end{align}
with the shorthand notation, $\int_{\bar{\zeta},\zeta}=\int\int{\rm d}\bar{\zeta}{\rm d}\zeta{\rm e}^{-\bar{\zeta}\zeta}$.
Thanks to these decompositions, we can independently carry out the Grassmann integration over $\chi$ and $\bar{\chi}$ at each lattice site as follows,
\begin{align}
    &\mathcal{T}^{(f)}_{
        \zeta_{1}\xi_{1}
        \zeta_{2}\xi_{2}
        \bar{\xi}_{1}\bar{\zeta}_{1}
        \bar{\xi}_{2}\bar{\zeta}_{2}
        ,a_{1}(n)a_{2}(n)
    }
    =\int{\rm d}\bar{\chi}{\rm d}\chi{\rm e}^{-m\bar{\chi}\chi}
    \prod_{\nu}
    {\rm e}^{
        \left(
        \eta_{\nu}(n){\rm e}^{{\rm i}\pi a_{\nu}(n)}\bar{\chi}\zeta_{\nu}
        +
        \chi\bar{\zeta}_{\nu}
        +
        \eta_{\nu}(n){\rm e}^{-{\rm i}\pi a_{\nu}(n)}\chi\xi_{\nu}
        +
        \bar{\chi}\bar{\xi}_{\nu}
        \right)
        /\sqrt{2}
    }
    .
\end{align}
Solving this integral, we obtain
\begin{align}
    &\mathcal{T}^{(f)}_{
        \zeta_{1}\xi_{1}
        \zeta_{2}\xi_{2}
        \bar{\xi}_{1}\bar{\zeta}_{1}
        \bar{\xi}_{2}\bar{\zeta}_{2}
        ,a_{1}(n)a_{2}(n)
    }
    =
    \left[
	\delta_{i_{1}+i_{2}+j'_{1}+j'_{2},1}
        \delta_{i'_{1}+i'_{2}+j_{1}+j_{2},1}
        +
        m\delta_{i_{1}+i_{2}+j'_{1}+j'_{2},0}
        \delta_{i'_{1}+i'_{2}+j_{1}+j_{2},0}
    \right]
    \nonumber\\
    &\times
    \left(\frac{1}{\sqrt{2}}\right)^{
        \sum_{\nu}(i_{\nu}+j_{\nu}+i'_{\nu}+j'_{\nu})
    }
    {\rm e}^{
	{\rm i}\pi\left[
	(i_{1}-j_{1})a_{1}(n)+(i_{2}-j_{2})a_{2}(n)
	\right]
    }
    (-1)^{
        j_{1}(i_{2}+j'_{1}+j'_{2})
        +
        j_{2}(j'_{1}+j'_{2})
        +
        i'_{1}j'_{2}
        +
        n_{1}(i_{2}+j_{2})
    }
    .
\end{align}

\section{Derivation of the correlation length}
\label{app:mass_gap}
In this appendix, we discuss the derivation of the correlation length employed in section \ref{sec:mass_gap}.

First, we review the method of estimating the correlation length based on the entanglement entropy, which is adopted in Ref.~\cite{Dempsey:2023gib}.
On an $N$-site chain with lattice spacing $a$, when the system is at the criticality, the entanglement entropy for a subsystem of the leftmost $xN$ sites, $S_{x}(N,a)$, is given by
\begin{align}
\label{eq:entanglement_entropy}
    S_{x}(N,a)=
    \frac{c}{6}\log\left(
        \frac{2N}{\pi}\sin\pi x
    \right)
    +
    {\rm const.},
\end{align}
with the central charge $c$~\cite{Calabrese:2009qy}.
When the system is not exactly massless, the central charge $c$ takes nonzero values for small $N$ but starts to vanish around $N\sim\xi/a$.
In other words, investigating the central charge $c(N)$ as a function of $N$, one can estimate the correlation length up to a multiplicative constant $b$ by $b\xi/a \sim N'$, where $N'$ is the smallest system size satisfying $c(N') \approx 0$.

In our study, we follow a similar procedure to determine the correlation length, excepting that we extract the central charge $c$ from the largest eigenvalue of the transfer matrix $\lambda_{0}$ via
\begin{align}
    \lambda_{0}
    =
    {\rm e}^{-f_{\infty}V+\pi c/6},
    \label{eq:lambda_0}
\end{align}
where $f_{\infty}$ is a nonuniversal constant. 
There is an established way to extract $c$ within the TRG algorithm that is proposed in Ref.~\cite{PhysRevB.80.155131}.
We follow this way within the framework of the Grassmann tensor network formulation. 
Note that Eq.~\eqref{eq:Grassmann_TM} gives the transfer matrix.
The calculation of $\lambda_{0}$ on a different volume is computationally reasonable in the TRG approach because one can calculate it at each iteration step. However, the volume is limited to the power of $2$.

We calculate the largest eigenvalue of the transfer matrix at each iteration step in TRG. Using Eq.~(\ref{eq:lambda_0}), we obtain the central charge at the $i$-th iteration step as
\begin{align}
    c(V_i)
    =
    -\frac{6V_{i+1}}{\pi}\left\{\frac{\log_2\lambda_0(i+1)}{V_{i+1}}-\frac{\log_2\lambda_0(i)}{V_{i}} \right\} \ ,
    \label{eq:c}
\end{align}
where $V_i=2^i$ is the volume and $\lambda_0(i)$ is the largest eigenvalue at the $i$-th iteration step.

\begin{figure}[htbp]
    \centering
    \includegraphics[width=0.98\hsize]{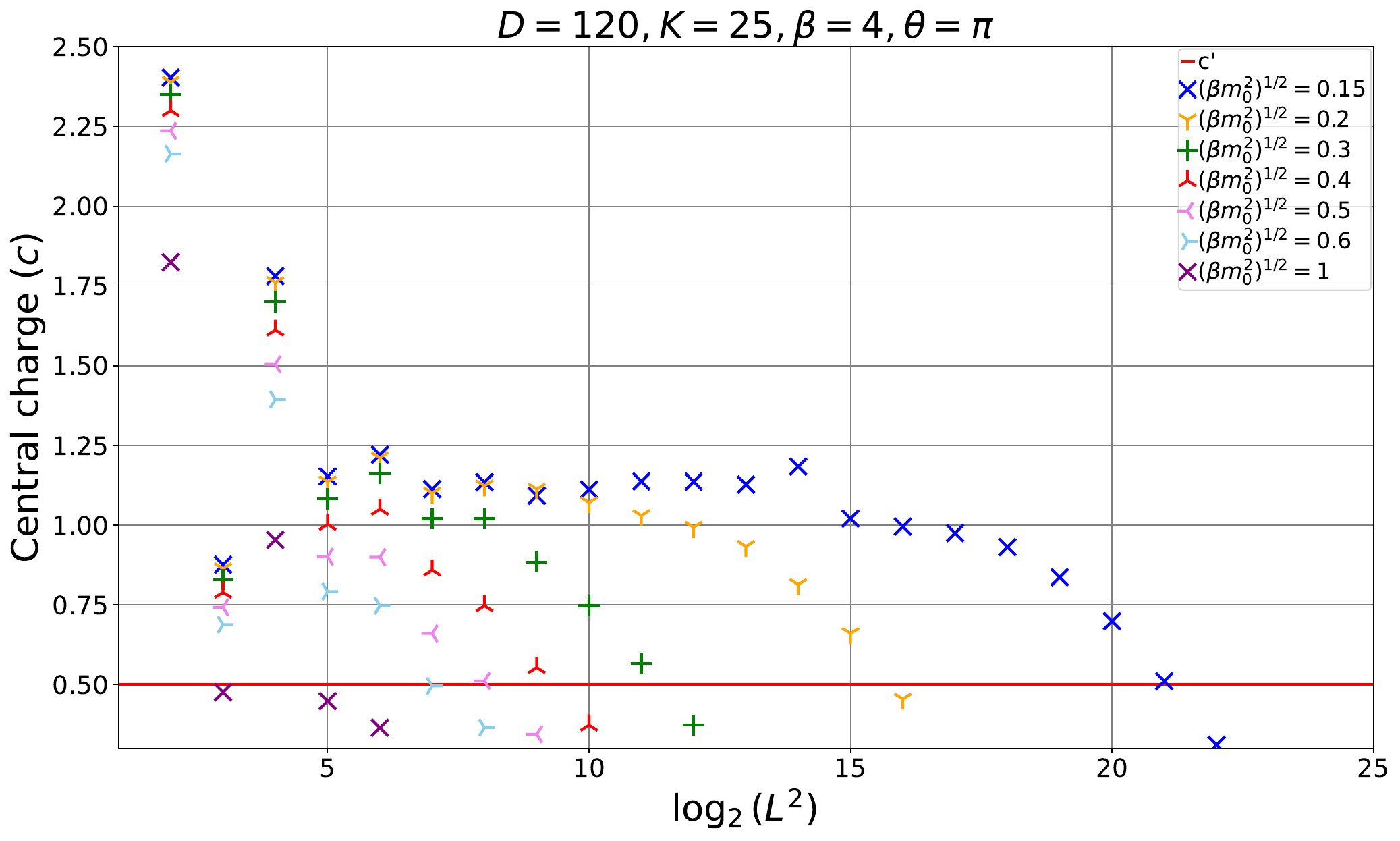}
    \caption{
        Central charge as a function of volume.
    }
    \label{fig:c_charge}
\end{figure}

Fig.~\ref{fig:c_charge} shows the results of $c(V)$ at $\theta=\pi$ and $\beta=4$ as a function of volume $V$. 
In Fig.~\ref{fig:c_charge}, $c$ lies around $1$ in the small volume, whereas it starts to deviate from $1$ in the large volume. 
The value of $c$ around $1$ continues at a larger volume for a smaller mass $m_{0}$.
This behavior may be explained by the finite-volume effect associated with the $c=1$ CFT due to a smaller system size than the correlation length $\xi/a$. 
This value of $c=1$ is consistent with that of the theoretical prediction of the massless theory, $SU(2)_1$ WZW model. At the large volume where the two-fold degeneracy appears, the system is expected to develop a gap, and the central charge should vanish.
Therefore, the correlation length can be estimated by the (square root of) the smallest volume $V'$ where $c(V') \ll 1$.
In actual calculation, we impose the threshold $c'=0.5$ as the boundary between the region for $c(V) \approx 1$ and $c(V') \ll 1$, and determine $V'$ as the smallest volume that satisfies $c(V') < c'$.
In the right figure of Fig.~\ref{fig:degeneracy_bm10000}, the plateau of two-fold degeneracy can be seen only for $\log_2 (L^2)\gtrsim 17$ at $\theta=\pi$.
In the smaller volume, the degeneracy cannot be integer-valued because of the finite-volume effect. We confirmed that the volume range where this finite-volume effect appears is consistent with the range where $V\lesssim V'$ is realized in Fig.~\ref{fig:c_charge}.

To obtain the precise value of $V'$, we pick up two points in the vicinity of $c'$, interpolate them by a linear function, and identify the intersection of the linear function with $c'$. 
We then estimate the correlation length with a multiplicative constant $b$ by $b\xi/a \simeq \sqrt{V'}$. 
The values of $\log_2 V' \simeq 2\log(b\xi/a)$ obtained in the above procedure are plotted in Fig.~\ref{fig:mass_gap}.
We have also confirmed that the slope in Fig.~\ref{fig:mass_gap} is stable under the change of the threshold $c'$ around $c'=0.5$.

\bibliographystyle{utphys.bst}
\bibliography{bib/kanno,bib/trg_algorithm,bib/trg_formulation,bib/trg_gauge,bib/trg_grassmann,bib/trg_review,bib/conti_dof}

\end{document}